\documentclass[]{pasj01}

\Received{}
\Accepted{}
 
\usepackage{natbib}
\usepackage{url}
\usepackage{comment}
\usepackage{threeparttable}

\newcommand{\HI}{\mbox{H\,{\sc i}}}
\newcommand{\HII}{\mbox{H\,{\sc ii}}}

\newcommand{\sigsta}{$\sigma_{{\rm ^{12}CO}}$}
\newcommand{\sigstai}{$\sigma_{{\rm ^{12}CO},i}$}

\begin{document} 

\title{ 
CO Multi-line Imaging of Nearby Galaxies (COMING). X. Physical conditions of molecular gas and the local SFR-Mass relation}

\author{Kana \textsc{Morokuma-Matsui}\altaffilmark{1,8}}%
\altaffiltext{1}{Institute of Astronomy, Graduate School of Science, The University of Tokyo, 2--21--1 Osawa, Mitaka, Tokyo 181--0015, Japan}
\email{kanamoro@ioa.s.u-tokyo.ac.jp}

\author{Kazuo \textsc{Sorai}\altaffilmark{2,3,4,5}}
\altaffiltext{2}{Department of Physics, Faculty of Science, Hokkaido University, Kita 10 Nishi 8, Kita-ku, Sapporo 060-0810, Japan}
\altaffiltext{3}{Department of Cosmosciences, Graduate School of Science, Hokkaido University, Kita 10 Nishi 8, Kita-ku, Sapporo 060--0810, Japan}

\author{Yuya \textsc{Sato}\altaffilmark{4}}
\author{Nario \textsc{Kuno}\altaffilmark{4,5}}
\altaffiltext{4}{Department of Physics, Graduate School of Pure and Applied Sciences, University of Tsukuba, 1--1--1 Tennodai, Tsukuba, Ibaraki 305--8577, Japan}
\altaffiltext{5}{Tomonaga Center for the History of the Universe, University of Tsukuba, 1--1--1 Tennodai, Tsukuba, Ibaraki 305--8571, Japan}

\author{Tsutomu T. \textsc{Takeuchi}\altaffilmark{6,7}}
\altaffiltext{6}{Division of Particle and Astrophysical Science, Nagoya University, Furo-cho, Chikusa-ku, Nagoya 464--8602, Japan}
\altaffiltext{7}{The Research Center for Statistical Machine Learning, the Institute of Statistical Mathematics, 10--3 Midori-cho, Tachikawa, Tokyo 190--8562, Japan}

\author{Dragan \textsc{Salak}\altaffilmark{5}}

\author{Yusuke \textsc{Miyamoto}\altaffilmark{8}}
\altaffiltext{8}{National Astronomical Observatory of Japan, National Institutes of Natural Sciences, 2--21--1 Osawa, Mitaka, Tokyo 181--8588, Japan}

\author{Yoshiyuki \textsc{Yajima}\altaffilmark{3}}

\author{Kazuyuki \textsc{Muraoka}\altaffilmark{9}}
\altaffiltext{9}{Department of Physical Science, Graduate School of Science, Osaka Prefecture University, 1--1 Gakuen-cho, Naka-ku, Sakai, Osaka 599--8531, Japan}

\author{Hiroyuki \textsc{Kaneko}\altaffilmark{8,10}}
\altaffiltext{10}{Joetsu University of Education, 1, Yamayashiki-machi, Joetsu, Niigata 943--8512, Japan}



\maketitle

\begin{abstract}

We investigate the molecular gas properties of galaxies across the main sequence of star-forming (SF) galaxies in the local Universe using $^{12}$CO($J=1-0$) (hereafter $^{12}$CO) and $^{13}$CO($J=1-0$) ($^{13}$CO) mapping data of 147 nearby galaxies obtained in the COMING project, a legacy project of the Nobeyama Radio Observatory.
In order to improve signal-to-noise ratios of both lines, we stack all the pixels where $^{12}$CO emission is detected after aligning the line center expected from the first-moment map of $^{12}$CO.
As a result, $^{13}$CO emission is successfully detected in 80 galaxies with a signal-to-noise ratio larger than three.
The error-weighted mean of integrated-intensity ratio of $^{12}$CO to $^{13}$CO lines ($R_{1213}$) of the 80 galaxies is $10.9$ with a standard deviation of $7.0$.
We find that
(1) $R_{1213}$ positively correlates to specific star-formation rate (sSFR) with a correlation coefficient of $0.46$, and
(2) both flux ratio of IRAS 60~$\mu$m to 100~$\mu$m ($f_{60}/f_{100}$) and inclination-corrected linewidth of $^{12}$CO stacked spectra (\sigstai) also correlate with sSFR for galaxies with the $R_{1213}$ measurement.
Our results support the scenario where $R_{1213}$ variation is mainly caused by the changes in molecular-gas properties such as temperature and turbulence.
The consequent variation of CO-to-H$_2$ conversion factor across the SF main sequence is not large enough to completely extinguish the known correlations between sSFR and $M_{\rm mol}/M_{\rm star}$ ($\mu_{\rm mol}$) or star-formation efficiency (SFE) reported in previous studies, while this variation would strengthen (weaken) the sSFR-SFE (sSFR-$\mu_{\rm mol}$) correlation.
\end{abstract}

\section{Introduction}

Galaxies show distinctive distributions on the plot of stellar mass ($M_{\rm star}$) and star-formation rate (SFR): 
some galaxies are on the sequence showing a positive correlation between the two values, and some locate far below the correlation in the SFR direction.
The former are called the main sequence of star-forming (SF) galaxies \citep[hereafter SF main sequence, e.g.,][]{Noeske:2007lr,Elbaz:2007ub,Daddi:2007un} and the latter are called quenched or passive galaxies.
Investigating how the galaxy properties change along and across the SF main sequence leads to essential understanding of galaxy evolution \citep[e.g.,][]{Wuyts:2011uq}.
Provided that stars are formed in molecular clouds in the local Universe, it is vital to study the molecular gas properties along and across the SF main sequence.

It is reported that there are positive correlations between specific SFR (sSFR $=$ SFR/$M_{\rm star}$, or an offset from the SF main sequence, $\Delta$(MS)) with molecular gas mass ($M_{\rm mol}$) to $M_{\rm star}$ ratio ($\mu_{\rm mol}=M_{\rm mol}/M_{\rm star}$) and star-formation efficiency (SFE $=$ SFR/$M_{\rm mol}$), i.e., galaxies with higher sSFR tend to have more abundant molecular gas and form stars more efficiently from the molecular gas \citep[e.g.,][]{Saintonge:2012nj,Genzel:2015gn,Scoville:2017jw,Tacconi:2018pv}.
These correlations still hold for spatially resolved data (\citealt{Lin:2017ee,Ellison:2020dm,Ellison:2020kf}; Kajikawa in prep.).
In addition, it is known that these correlations are irrespective of the environment \citep{Koyama:2017lf} or mass concentration of stellar components of galaxies \citep{Koyama:2019vy}, indicating their essentiality for galaxy evolution.
Furthermore, recent studies claimed that the scatter of the SF main sequence is primarily due to the SFE variety with a secondary role of $\mu_{\rm mol}$ using spatially-resolved molecular-gas data and optical integral-field-unit data \citep{Ellison:2020kf}.

Molecular rotational transitions have been used to explore the physical properties of molecular gas in galaxies.
Especially, the lowest transition lines ($J=1-0$) of carbon monoxide, $^{12}$CO and its isotopes are ideal tracers of cold ($\sim10$~K) molecular gas which is raw material of star formation, since the corresponding energy gaps ($h \nu_{\rm rest}$ where $h$ is the Planck constant and $\nu_{\rm rest}$ is rest frequency of the line) divided by the Boltzmann constant are $\sim5.53$~K for $^{12}$CO($J$=1-0) ($\nu_{\rm rest}=115.271$~GHz, hereafter $^{12}$CO) and $\sim5.29$~K for $^{13}$CO($J$=1-0) ($\nu_{\rm rest}=110.201$~GHz, hereafter $^{13}$CO), respectively.
$^{12}$CO emission is generally optically thick but strong even in extragalactic objects, thus it is widely used to measure the molecular gas mass of galaxies by assuming CO-to-H$_2$ conversion factors \citep[e.g.,][and references therein]{Rickard:1975if,Solomon:1975jr,Young:1982sd,Bolatto:2013vn}.
On the other hand, $^{13}$CO is used as an optically thin tracer of molecular gas and known to be well correlated with dust extinction in molecular clouds in the Milky Way \citep[e.g.,][]{Dickman:1978nn,Frerking:1982zq,Lada:1994ya}.

Detailed studies of the line ratio of $^{12}$CO/$^{13}$CO (hereafter $R_{1213}$) have revealed a wide variety of $R_{1213}$ within the Milky Way.
In a giant molecular cloud (GMC) where stars are formed, $R_{1213}$ is reported to be $\sim3$ at the center and $\sim5$ for the whole cloud \citep{Gordon:1976ly,Solomon:1979zd,Polk:1988ly}, which are much smaller than the abundance ratio of [$^{12}$CO]/[$^{13}$CO] inferred from the $^{12}$C/$^{13}$C ratio which has a radial gradient, where $\sim20$ at the galaxy center and $\sim60$ in the Solar neighbourhood \citep{Milam:2005yb,Halfen:2017sd}.
This suggests that bulk of $^{12}$CO emission is optically thick in GMCs.
On the other hand, larger $R_{1213}$ values are reported in the peripheries of GMCs \citep[$R_{1213}\sim20$,][]{Sakamoto:1994di}, high-latitude molecular clouds \citep[$\sim10.5$,][]{Blitz:1984ky}, and small molecular clouds with a size of $<1.0$~pc \citep[$\sim40$,][]{Knapp:1988gf}, suggesting lower optical depth of $^{12}$CO line in those components than in GMCs.
\cite{Polk:1988ly} obtained a relatively high $R_{1213}$ value of $\sim7$ when they calculate an average $R_{1213}$ value of seven $0.5^\circ \times 0.5^\circ$ regions in the Galactic plane and suggested a non-negligible contribution of the lower optical depth (diffuse) molecular gas to large-scale $^{12}$CO emission of the Milky Way.

Since the early studies of CO isotopes of local galaxies in 1970s (\citealt{Encrenaz:1979dw}, upper limits by \citealt{Rickard:1977ox}), various and fairly larger $R_{1213}$ values than those of GMCs have been reported in nearby galaxies, for example, a typical $R_{1213}$ of $\sim10$ in spiral galaxies \citep[e.g.,][]{Young:1982sd,Young:1986hh,Weliachew:1988tx,Sandqvist:1988fh,Sage:1991rj,Wright:1993ic,Braine:1993iw,Aalto:1994yg,Xie:1994do,Matsushita:1998si,Paglione:2001if,Hirota:2010af,Watanabe:2011na,Davis:2014tq,Vila-Vilaro:2015ip,Morokuma-Matsui:2015cr,Cormier:2018lt,Lee:2018rt}, $\sim10$ in early-type galaxies \citep[ETGs, e.g.,][]{Eckart:1990rp,Sage:1990yz,Wild:1997dm,Crocker:2012vs,Davis:2014tq,Alatalo:2015ca}, $\sim15$ in irregular galaxies \citep[e.g.,][]{Becker:1991da}, $\sim30$ in merging galaxies \citep[e.g.,][]{Aalto:1991bn,Casoli:1992zy,Aalto:1997wk,Taniguchi:1998yi,Aalto:2010yx}, starburst galaxies \citep[e.g.,][]{Stark:1984av,Young:1984sl,Kikumoto:1998xw}, and infrared-bright galaxies \citep[e.g.,][]{Garay:1993wo,Aalto:1995ir,Cao:2017bf,Sliwa:2017aq,Herrero-Illana:2019lc}, $\sim10$ in Seyfert galaxies \citep[e.g.,][]{Papadopoulos:1998fh,Papadopoulos:1999wa}, and $\sim30$ in $z\sim3$ galaxies with $J>1$ transitions \citep[e.g.,][]{Henkel:2010av,Danielson:2013yl,Spilker:2014lc,Bethermin:2018jq}.
The large $R_{1213}$ values would reflect a presence of substantial amount of low-optical depth (diffuse) components in those galaxies \citep[e.g.,][]{Polk:1988ly,Aalto:1995ir,Garcia-Burillo:1992as,Sakamoto:1997gr,Hirota:2010af,Morokuma-Matsui:2015cr}.
These studies are suggestive of a dependence of $R_{1213}$ on SF activities in galaxies but the relationship between $R_{1213}$ and sSFR or $\Delta$(MS) has not been systematically explored so far.

In this study, we compare the $R_{1213}$ and sSFR of galaxies targeted in the CO Multi-line Imaging of Nearby Galaxies project \citep[COMING,][]{Sorai:2019hs}, a legacy project making use of the 45-m radio telescope at the Nobeyama Radio Observatory\footnote{The Nobeyama 45-m radio telescope is operated by Nobeyama Radio Observatory, a branch of National Astronomical Observatory of Japan.} in order to investigate the molecular gas properties across the SF main sequence of galaxies.
The COMING project {\it simultaneously} obtained $^{12}$CO, $^{13}$CO, and C$^{18}$O($J=1-0$) mapping data of 147~galaxies with various sSFR located at the distances of $\lesssim40$~Mpc, which makes it one of the largest mapping  surveys of galaxies in multiple CO isotopes.
Provided that the $^{12}$CO and $^{13}$CO data in the COMING project are obtained in the same observing conditions including the accuracy of telescope pointings, $R_{1213}$ measurements do not suffer from the relevant uncertainties.

The structure of this paper is as follows:
we first briefly describe the data we used and data analysis, specifically stacking analysis, in section~\ref{sec:data_analysis},
we present the basic properties and the relationship between $R_{1213}$ and sSFR of our sample galaxies in section~\ref{sec:results},
based on the results, we discuss the origin of the $R_{1213}$ variation and its implication for the consequent variety in CO-to-H$_2$ conversion factor ($\alpha_{\rm CO}$) in local galaxies as well as high-redshift galaxies in section~\ref{sec:discussions},
and then we summarize this study in section~\ref{sec:summary}.
Throughout this study, we assume the Kroupa initial mass function (IMF) \citep{Kroupa:2001xf,Kroupa:2003do}.
For a correlation check, we refer to three values, Pearson's correlation coefficient $r$, Spearman's rank correlation coefficient $\rho$, and Kendall's rank correlation coefficient $\tau$.
The first one is a parametric measure and the latter two are non-parametric measures.
Although the obtained values are different from method to method, overall trends are almost consistent in the three methods.
We basically adopt the widely-used Spearman's $\rho$ to discuss the correlation of two galaxy parameters in this paper.

\section{Data \& Analysis}
\label{sec:data_analysis}

\begin{figure*}[h]
\begin{center}
\includegraphics[width=\textwidth, bb=0 0 1431 502]{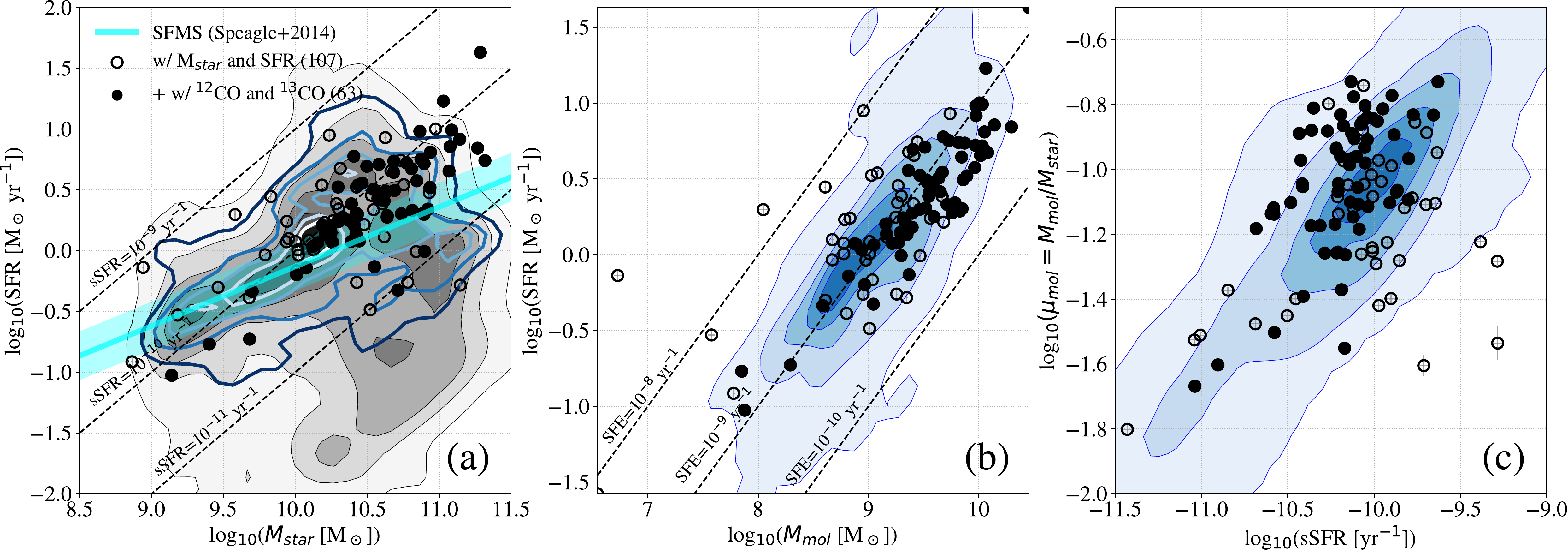}
\end{center}
\caption{
Basic properties of the COMING galaxies:
(a) $M_{\rm star}$-SFR relation,
(b) $M_{\rm mol}$-SFR relation (``Kennicutt-Schmidt relation''), and 
(c) sSFR-$\mu_{\rm mol}$ relation.
COMING galaxies are indicated as circle symbols, and among them, those with $R_{1213}$ measurements are indicated as filled black circles.
In the panel (a), the grey-filled contour indicates galaxies observed in the xCOLD GASS project and the thick contours with different shades of blue indicates xCOLD GASS galaxies with CO detection \citep{Saintonge:2011hl,Saintonge:2017ve}.
The SF main sequence at $z=0$ is indicated as cyan solid line and the 0.2-dex scatter range is indicated as cyan shade \citep{Speagle:2014by}.
In the panels (b) and (c), the blue-filled contour indicates xCOLD GASS galaxies with CO detection, which are same galaxies indicated as thick contours in the panel (a).
}
\label{fig:basic}
\end{figure*}

In this study, we used physical parameters of the COMING galaxies such as $M_{\rm star}$, SFR, sSFR, $M_{\rm mol}$, SFE, $\mu_{\rm mol}$, $R_{1213}$, and velocity dispersion of stacked {\bf $^{12}$CO} spectra.
Here, we briefly mention the derivation of these parameters from the observed data in this section.

\subsection{Molecular lines: COMING $^{12}$CO and $^{13}$CO data}
\label{sec:mmol}

We used $^{12}$CO and $^{13}$CO data obtained in the COMING project \citep{Sorai:2019hs} to calculate the integrated intensity ratio of:
\begin{equation}
R_{1213} = I_{\rm ^{12}CO}/I_{\rm ^{13}CO},
\end{equation}
where $I_{\rm ^{12}CO}$ and $I_{\rm ^{13}CO}$ are integrated intensities in units of K~km~s$^{-1}$ of $^{12}$CO and $^{13}$CO emission lines, respectively.
The angular and velocity resolutions of the COMING data are $\sim17$ arcsec ($\sim1.6$~kpc at 20~Mpc) and 10~km~s$^{-1}$, respectively.
We stacked all the $^{12}$CO and $^{13}$CO spectra of pixels where $^{12}$CO emission is detected with a signal-to-noise ratio of larger than three for each galaxy after aligning the line center expected from the first-moment map of $^{12}$CO line \citep[hereafter, ``VA'' stacking,][]{Schruba:2011zr,Caldu-Primo:2013lb,Morokuma-Matsui:2015cr}.
The number of stacked pixels differs from galaxy to galaxy and ranges from 20 to 2431 (the median, first and third quartiles are 274.5, 154.5, and 447.5, respectively)\footnote{Since the number of galaxies with $^{13}$CO detection is even (80), the median, first and third quartiles are calculated as means of the central two values of the whole sample, the first quarter sample, and the last quarter sample, respectively when the galaxies are sorted according to the number of the stacked pixels.}.
With the VA stacking analysis, $^{13}$CO emission line is detected in 80 galaxies with a signal-to-noise ratio of $>3$.
It should be noted that the number of galaxies with a ``secure'' detection of $^{13}$CO emission in the original data (more than three pixels with $\geq4$ sigma detection in the $^{13}$CO integrated intensity and $\geq3$ sigma detection in the $^{12}$CO integrated intensity) is only 52 \citep{Sorai:2019hs}.

Velocity dispersions of stacked $^{12}$CO spectra, \sigsta~are derived by the single Gaussian fitting.
Since this value encompasses both random motion among clouds and/or volume-filling gas and velocity gradient due to galactic rotation within the $\sim1$~kpc beam, we correct the effect of inclination ($i$) by dividing the value by sin($i$) as \sigstai=\sigsta/sin($i$).
$^{12}$CO data is also used to calculate molecular gas mass of galaxies.
We adopt $M_{\rm mol}$ from \cite{Sorai:2019hs} which assumed the standard CO-to-H$_2$ conversion factor of the Milky Way of $2.0\times10^{20}$~cm$^{-2}$~(K km s$^{-1}$)$^{-1}$ \citep{Bolatto:2013vn}.

\subsection{Stellar mass and SFR: WISE and GALEX}
\label{sec:mstarsfr}

We adopt $M_{\rm star}$ values derived in \cite{Sorai:2019hs}, which used $3.4$~$\mu$m data obtained with the Wide-field Infrared Survey Explorer \citep[WISE,][]{Wright:2010oi} and an empirical relation between $3.4$~$\mu$m luminosity and $M_{\rm star}$ presented in \cite{Wen:2013wt} assuming the Kroupa IMF.
\cite{Sorai:2019hs} reported that their $M_{\rm star}$ values are $\sim20$~\% lower than those calculated in the Spitzer Survey of Stellar Structure in Galaxies project \citep[S$^4$G][]{Sheth:2010mz} due to their conservative subtractions of foreground stars.
SFR map of the COMING galaxies is generated using WISE $22$~$\mu$m and far-ultraviolet (FUV) imaging data obtained with the Galaxy Evolution Explorer \citep[GALEX,][]{Martin:2005wd} as
\begin{equation}
\Sigma_{\rm SFR} = 3.2\times10^{-3} \times I_{22\mu m} + 8.1 \times 10^{-2} \times I_{\rm FUV},
\end{equation}
where $\Sigma_{\rm SFR}$ is SFR surface densities in units of $M_\odot$~yr$^{-1}$~kpc$^{-2}$ and $ I_{22\mu m}$ and $I_{\rm FUV}$ are WISE/$22\mu$m and GALEX/FUV intensities, respectively, in units of MJy~sr$^{-1}$ \citep{Casasola:2017zv}.
The SFR calibration method adopted in \cite{Casasola:2017zv} assumes the default IMF in {\tt STATBURST99} \citep{Leitherer:1999ye}.
To convert from the SFR with the {\tt STATBURST99} IMF to SFR with the Salpeter IMF \citep{Salpeter:1955bz}, one should multiply the value by a factor of 1.59.
Furthermore, to convert from the SFR with Salpeter IMF to SFR with Kroupa IMF, one should multiply the value by a factor of 0.67.
Details in estimating SFR and star-formation properties will be presented in the forthcoming paper (Takeuchi et al. in prep.).
In this study, we limit our sample to the COMING galaxies with both the WISE $22$~$\mu$m and the GALEX FUV detections, resulting in 107 galaxies.
The number of galaxies with $M_{\rm star}$ and SFR measurements and $^{13}$CO detections reduces to $63$.
The $63$ galaxies consist of spiral galaxies \citep[S, SA, SAB, or SB classifications in the Third Reference Catalogue of Bright Galaxies, RC3,][]{de-Vaucouleurs:1991pr} and no interacting galaxies, irregular galaxies, nor ETGs are included.

\subsection{Dust emission: IRAS $60~\mu$m and $100~\mu$m data}
\label{sec:iras}
We crossmatch the 147 COMING galaxies with the IRAS Revised Bright Galaxy Sample \citep[RBGS,][]{Sanders:2003ci}, which lists total fluxes covering extended emission of galaxies in IRAS bands.
The IRAS RBGS consists of 629 galaxies with $60~\mu$m flux of $>5.24$~Jy.
The matching results in 111 galaxies, and 62 galaxies have all the measurements of $M_{\rm star}$, SFR, stacked $^{12}$CO and $^{13}$CO emissions, and IRAS fluxes at $60~\mu$m and $100~\mu$m.

\section{Results}
\label{sec:results}

\begin{figure*}[h]
\begin{center}
\includegraphics[width=\textwidth, bb=0 0 1397 500]{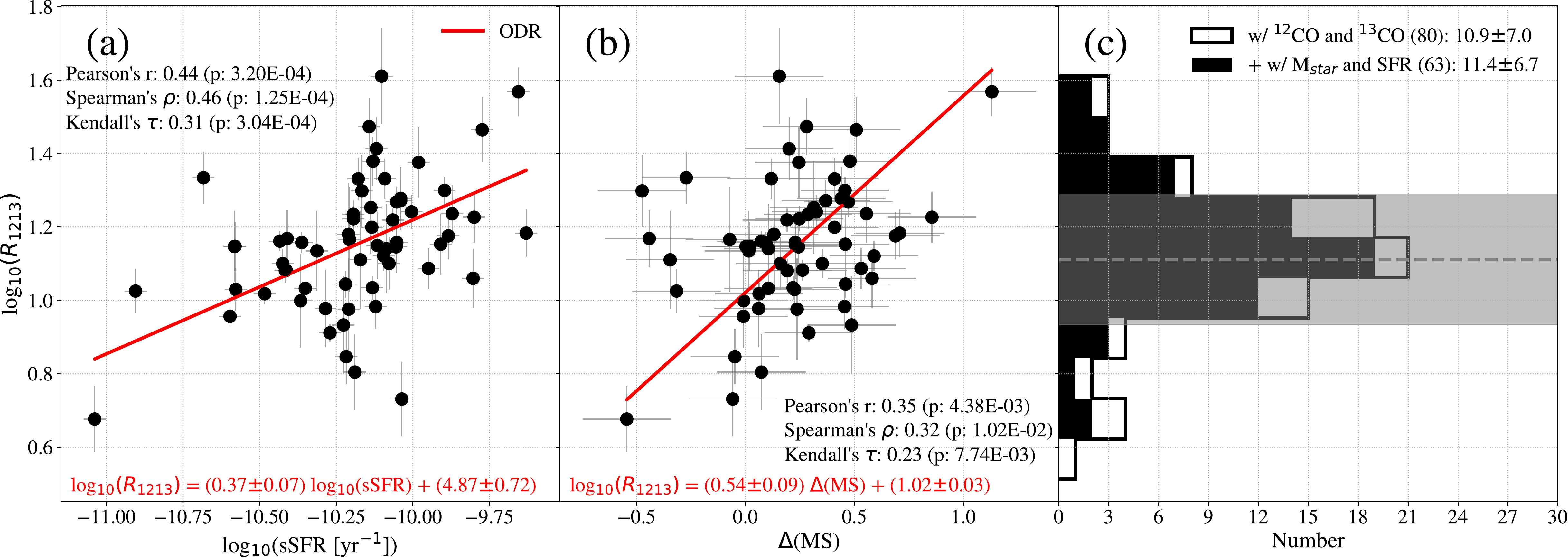}
\end{center}
\caption{
The obtained relations between (a) $R_{1213}$ and sSFR and (b) $R_{1213}$ and $\Delta$(MS) of the 63 COMING galaxies.
Panel (c) shows the histogram of the $R_{1213}$ values for the 80 galaxies with $R_{1213}$ measurements and the 63 galaxies with both $R_{1213}$ and sSFR measurements.
The fitting functions with ODR are presented at the bottom of the panels (a) and (b).
Grey-dashed line and grey-shaded regions in panel (c) indicates an error-weighted mean and a standard deviation of the 63 COMING galaxies.
}
\label{fig:r1213_ssfr}
\end{figure*}

In figure~\ref{fig:basic}, we present basic properties of the COMING galaxies and compare them with galaxies in the extended CO Legacy Database for GASS project \citep[xCOLD GASS,][]{Saintonge:2011hl,Saintonge:2017ve}, which is the largest $^{12}$CO survey targeting 532 galaxies at $0.01<z<0.05$.
The SF main sequence at the local Universe is also indicated as a solid cyan line as a reference in the panel (a) \citep{Speagle:2014by}.
We can see that the most COMING sample is above or on the SF main sequence and some are located on the so-called ``green valley'' regime.
It should be noted that our sSFR estimations of the COMING galaxies are expected to be higher than those in literatures since the adopted $M_{\rm star}$ values are known to be $\sim20$~\% lower than the previous studies as mentioned in section~\ref{sec:mstarsfr}.
Overall, the COMING and the CO-detected xCOLD-GASS galaxies share similar star-formation properties such as sSFR, SFE and $\mu_{\rm mol}$.
In panel (c) of the same figure, it is also seen that galaxies without $^{13}$CO detection indicated as open black circles tend to have low $\mu_{\rm mol}$ and/or high sSFR values.

An error-weighted mean and a standard deviation of $R_{1213}$ for the 80 COMING galaxies with $^{12}$CO and $^{13}$CO measurements are $10.9$ and $7.0$, respectively.
In case we limit sample to the 63 galaxies that are plotted on the figure~\ref{fig:r1213_ssfr}a, these value becomes $11.4$ (mean) and $6.7$ (standard deviation).
This is consistent with previous studies on $R_{1213}$ of local star-forming galaxies \citep[$\sim10-15$, e.g.,][]{Young:1986hh,Paglione:2001if,Vila-Vilaro:2015ip}.
Using $R_{1213}$ and sSFR of the 63 galaxies, we find a moderate correlation between these two values with a Spearman rank correlation coefficient $\rho$ of $0.46$ (see figure~\ref{fig:r1213_ssfr}a).
The possibility of a fake correlation for the $R_{1213}$-sSFR relation is discussed in the Appendix section.
The fitting function by the orthogonal distance regression (ODR) is shown in this figure.

We also examine the relationship between $R_{1213}$ and the SFR offset from the SF main sequence, $\Delta$(MS) that is calculated as
\begin{equation}
\Delta({\rm MS}) = \log({\rm SFR}) - \log({\rm SFR}_{\rm MS}(M_{\rm star}, t)),
\end{equation}
where SFR$_{\rm MS}$($M_{\rm star}$, $t$) is SFR of the SF main-sequence galaxies with mass of $M_{\rm star}$ [$M_\odot$] at the age of the Universe of $t$ [Gyr], which is calculated as 
\begin{equation}
\log({\rm SFR}_{\rm MS}(M_{\rm star}, t)=(0.84-0.026) \log(M_{\rm star}) - (6.51-0.11t)
\end{equation}
\citep{Speagle:2014by}\footnote{\cite{Speagle:2014by} compiled 64 observations of main-sequence galaxies at $0<z<6$ from 25 studies and derived the SF main sequence as a function of redshift after converting them to the same calibrations.}.
Here we adopt the same cosmology parameters as \cite{Speagle:2014by} of $(h, \Omega_{M}, \Omega_{\Lambda})=(0.7, 0.3, 0.7)$ to calculate the age of the Universe at $z=0$.
$R_{1213}$ seems to similarly but rather weakly correlate with $\Delta$(MS) compared to sSFR.
It should be noted that this tendency does not change even if another widely-used definition of the SF main sequence \citep{Whitaker:2012ok} is adopted, where the obtained Pearson's $r$, Spearman's $\rho$, and Kendall's $\tau$ are 0.41, 0.39, and 0.27, respectively.
Since $\Delta$(MS) depends on the definition of the main sequence of SF galaxies, we prefer to use sSFR, a more assumption free value, rather than $\Delta$(MS) to see possible variations in physical properties of molecular gas {\it across} the main sequence of SF galaxies in the following sections.

\section{Discussions}
\label{sec:discussions}

\begin{figure*}[h]
\begin{center}
\includegraphics[width=\textwidth, bb=0 0 2807 2821]{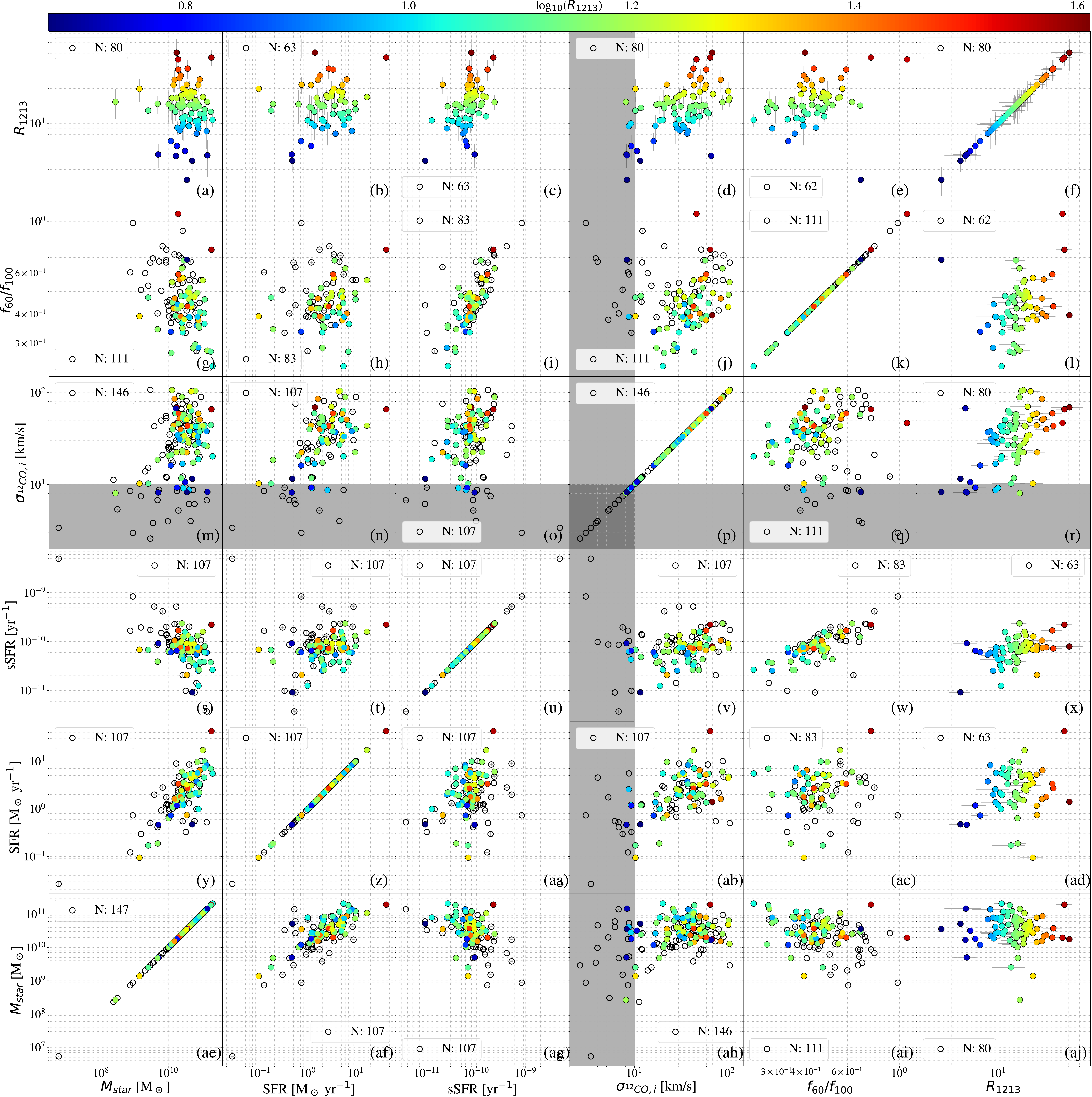}
\end{center}
\caption{
Correlation plots among six key parameters (from top to bottom and right to left: $R_{1213}$, $f_{60}/f_{100}$, \sigstai, sSFR, SFR, and $M_{\rm star}$).
All the galaxies with measurements of both abscissa and ordinate values are indicated as open circle, and filled circle indicates galaxies with $R_{1213}$ and sSFR measurements among them.
The colors of the symbols indicate the $R_{1213}$ value.
The grey-shaded region indicates \sigstai~$<10$~km~s$^{-1}$ that is the velocity resolution of the COMING data.
The number of galaxies with measurements of both abscissa and ordinate values is also indicated in each panel.
}
\label{fig:comp}
\end{figure*}

We found that $R_{1213}$ somehow relates to sSFR (and $\Delta$(MS)) suggesting that physical properties of molecular gas traced by $^{12}$CO can be different across the main-sequence of SF galaxies.
Here, we investigate relationship of $R_{1213}$ with the other physical parameters of galaxies in section~\ref{sec:corrplots} and discuss causes of the $R_{1213}$ variety in sections~\ref{sec:causes} and \ref{sec:morph}, and its implications for variety in CO-to-H$_2$ conversion factor ($\alpha_{\rm CO}$) in section~\ref{sec:implication}.
We also discuss possible effects of the $\alpha_{\rm CO}$ variation across the SF main sequence of galaxies on the known relations of sSFR with SFE and $\mu_{\rm mol}$ in section~\ref{sec:galevo}.

\subsection{Correlation plots for other parameters}
\label{sec:corrplots}

\begin{figure*}[h]
\begin{center}
\includegraphics[width=\textwidth, bb=0 0 1685 519]{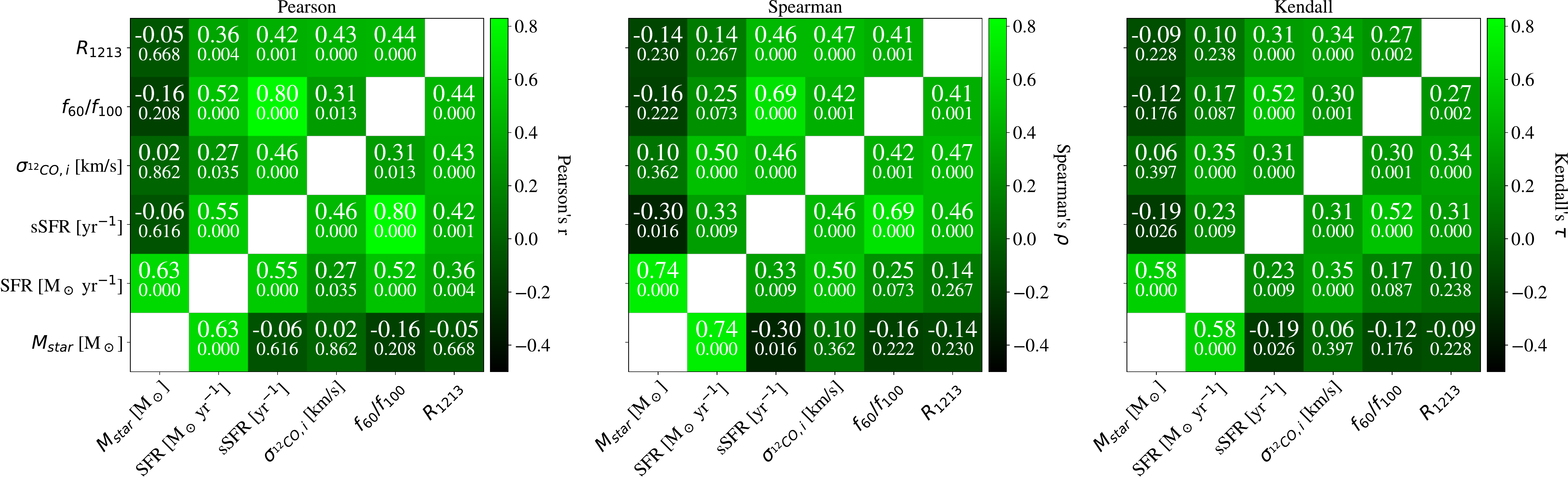}
\end{center}
\caption{
Correlation coefficients (upper row) and $p$-values (lower row) of the plots in each panel in figure~\ref{fig:comp} for galaxies with the $R_{1213}$ and sSFR measurements.
The correlation coefficients are measured in three difference methods:
(a) Pearson's correlation coefficient, $r$,
(b) Spearman's rank correlation coefficient, $\rho$,
(c) Kendall's $\tau$ coefficient.
The pixel color indicates the correlation coefficient value. 
Note that (a) is parametric and (b) and (c) are non-parametric methods.
(c) is normally more robust than (b) and generally (c)$<$(b) (see Appendix).
}
\label{fig:corr}
\end{figure*}

In order to investigate the $R_{1213}$-sSFR relation, we further compare $R_{1213}$ and sSFR with other four parameters including SFR, $M_{\rm star}$, \sigstai, and $f_{60}/f_{100}$ in figure~\ref{fig:comp}.
We focus on these values because SFR and $M_{\rm star}$ are respectively a numerator and a denominator of sSFR, and \sigstai and $f_{60}/f_{100}$ are measures of physical properties of molecular gas such as turbulence and temperature.
Figure~\ref{fig:corr} shows correlation coefficients of each relation shown in figure~\ref{fig:comp}, which are calculated for galaxies with $R_{1213}$ and sSFR measurements on each panel, i.e., the 63 galaxies indicated as filled circles (or the 62 galaxies for the panels that are related to $f_{60}/f_{100}$).

We find moderate correlations (correlation coefficients of $0.3-0.5$) between $R_{1213}$ and two parameters: \sigstai, and $f_{60}/f_{100}$ (figures~\ref{fig:comp} and \ref{fig:corr}).
Thus the $R_{1213}$ variation across the main sequence of SF galaxies is likely to be related to the varieties in physical properties of molecular gas.
Note that the correlation coefficients do not significantly change even if we simply use \sigsta~instead of \sigstai.
On the other hand, no correlation is found for the $R_{1213}$-$M_{\rm star}$ and the $R_{1213}$-SFR relations.
This suggests the correlation between $R_{1213}$ and sSFR is not a consequence of neither a positive correlation between $R_{1213}$ and SFR nor a negative correlation between $R_{1213}$ and $M_{\rm star}$.

The strongest correlation is found in the $f_{60}/f_{100}$-sSFR relation.
Considering that $f_{60}/f_{100}$ is a measure of dust temperature, this is consistent with previous studies showing higher dust temperature for galaxies with higher sSFR \citep[e.g.,][]{Magnelli:2014zr,Matsuki:2017nl}.
The stronger correlation of the $f_{60}/f_{100}$-sSFR relation than the one of the $R_{1213}$-sSFR relation suggests that the former is more fundamental than the latter.
Thus, we consider that the $R_{1213}$ correlates with sSFR primarily through the correlations of the sSFR-$f_{60}/f_{100}$ and $R_{1213}$-$f_{60}/f_{100}$ relations.

\sigstai~may also play a role in generating the $R_{1213}$-sSFR relation.
As aforementioned, $R_{1213}$ positively correlates with \sigstai.
In panels (o) or (v) of figure~\ref{fig:comp} (a comparison of \sigstai~and sSFR), we can see that the bulk of galaxies including those with $R_{1213}$ measurements are located on the upper left or lower right regions showing a moderate positive correlation with a Spearman's $\rho$ of $0.46$.
There also exist some outliers with \sigstai~$\lesssim10$~km~s$^{-1}$ showing a wide range of sSFR.
Consequently, the overall correlation becomes quite weak when including those outliers.
Note that these trends are still seen even when we used the stacked spectra of only disk region of galaxies.
It should be also noted that the \sigstai~values of the galaxies with \sigstai~$\lesssim10$~km~s$^{-1}$ are upper limits, considering the velocity resolution of $10$~km~s$^{-1}$ of the COMING data.

In summary, figures~\ref{fig:comp} and \ref{fig:corr} show that
(1) $R_{1213}$ correlates with \sigstai~and $f_{60}/f_{100}$,
(2) the correlation coefficients for the \sigstai-sSFR (0.46) and $f_{60}/f_{100}$-sSFR (0.69) relations are comparable to or higher than the one for the $R_{1213}$-sSFR relation (0.46), and 
(3) the correlation coefficient for the $f_{60}/f_{100}$-sSFR relation is higher than the one for the \sigstai-sSFR relation.
Therefore, the observed trend of $R_{1213}$ along sSFR is considered to be mainly caused by $f_{60}/f_{100}$ with a secondary role of \sigstai, while accurate measurements of \sigstai and $R_{1213}$ of those with \sigstai~$\lesssim10$~km~s$^{-1}$ are required for a definite conclusion.

\subsection{Causes of the $R_{1213}$ variety}
\label{sec:causes}

As a cause of the $R_{1213}$ variation, four candidates have been mainly considered since the early studies of $R_{1213}$ in nearby galaxies: changes in 
(1) physical conditions of molecular gas (density, temperature, and opacity),
(2) [$^{12}$CO]/[$^{13}$CO] abundance due to selective photodissociation,
(3) [$^{12}$CO]/[$^{13}$CO] abundance due to chemical fractionation, and
(4) $^{12}$C/$^{13}$C isotope abundance ratio due to stellar nucleosynthesis.
We consider that the scenario (1) is plausible for the observed $R_{1213}$ variation at least for our sample galaxies and the scenario (4) may play a secondary role. 
In this section, we explain each scenario and reasons why the scenarios (2)-(4) are not expected to play a major role.

$R_{1213}$ can be approximated as a function of opacity of $^{12}$CO ($\tau_{^{12}{\rm CO}}$) and the abundance ratio of [$^{12}$CO]/[$^{13}$CO] under the local thermodynamic equilibrium (LTE) assumption as\footnote{
Some studies approximates as $R_{1213}\propto \frac{1-\exp{(-\tau_{^{12}{\rm CO}})}}{1-\exp{(-\tau_{^{13}{\rm CO}})}} \sim \frac{1}{\tau_{^{13}{\rm CO}}}$ by assuming that $\tau_{\rm ^{12}CO}$ is large enough \citep{Paglione:2001if,Hirota:2010af,Cao:2017bf}.}:
\begin{equation}
R_{1213}\propto 
\frac{1-\exp{(-\tau_{^{12}{\rm CO}})}}{1-\exp{(-\tau_{^{13}{\rm CO}})}}
\sim \frac{1-\exp{(-\tau_{^{12}{\rm CO}})}}{1-\exp{\left (-\frac{\tau_{^{12}{\rm CO}}}{\rm [^{12}CO]/[^{13}CO]} \right)}},
\end{equation}
where $\tau_{^{13}{\rm CO}}$ is the optical depth of $^{13}$CO.
If $h\nu \ll k_{\rm B} T_{\rm k}$ (where $k_{\rm B}$, $\nu$, and $T_{\rm k}$ are the Boltzman constant, line frequency and kinetic temperature), $\tau_{^{12}{\rm CO}}$ scales as $\propto$~$\frac{N_{\rm ^{12}CO}}{\Delta v T^2_{\rm k}}$, where $N_{\rm ^{12}CO}$ and $\Delta v$ are $^{12}$CO column density and velocity dispersion, respectively.
Note that non-LTE analysis also predicts that $R_{1213}$ is a strong function of $\tau_{^{12}{\rm CO}}$ \citep{Cormier:2018lt}.
Thus, the $R_{1213}$ variation can be considered to reflect the $\tau_{^{12}{\rm CO}}$ variation as a first approximation, and $\tau_{^{12}{\rm CO}}$ is affected by temperature and turbulence of molecular gas.

Early studies have already noticed a correlation between $R_{1213}$ and $f_{60}/f_{100}$ (e.g., \citealt{Young:1986hh,Aalto:1995ir,Crocker:2012vs,Herrero-Illana:2019lc}, but see also \citealt{Vila-Vilaro:2015ip,Cao:2017bf}).
They discussed that high gas temperatures and turbulence (as traced by dust temperature) work against large optical depths in the $^{12}$CO.
However, no clear observational evidence has been found for a positive correlation between $R_{1213}$ with the velocity dispersion of $^{12}$CO line.
Although there are relevant studies to the $R_{1213}$-\sigstai~relation, such as the relationships of $R_{1213}$ with velocity dispersion of $^{13}$CO line \citep{Meier:2004cy} and with inclinations of galaxies \citep{Sage:1991rj}, both studies reported null results.
In this study, we re-examined and confirmed the correlation between $R_{1213}$ and $f_{60}/f_{100}$ with the COMING data and additionally found the correlation between $R_{1213}$ and \sigstai when we limit the sample that are used to investigate the $R_{1213}$-sSFR correlation.
These correlations also support the scenario (1), when we assume $\Delta v \sim $\sigstai, .

The abundance ratio of [$^{12}$CO]/[$^{13}$CO] can also affect $R_{1213}$.
The scenarios (2)-(4) are processes changing [$^{12}$CO]/[$^{13}$CO] ratio.
The selective photodissociation is expected to be dominant basically in low density ($n <10^2$~cm$^{-3}$) domains or in dense regions with very strong radiation fields \citep{Rollig:2013xw}.
UV radiation dissociates CO molecules and $^{13}$CO is less effective at self-shielding against UV photons than $^{12}$CO due to its lower abundance.
The chemical fractionation is expected to become important at deeper region of molecular clouds where the dust-shielding is effective for both $^{12}$CO and $^{13}$CO.
In this region with lower temperature of $<35$~K, $^{13}$CO is preferentially formed via the isotopic charge exchange reaction of:
\begin{equation}
^{12}{\rm CO} + ^{13}{\rm C}^+ \rightarrow ^{12}{\rm C}^+ + ^{13}{\rm CO} + \Delta {\rm E}.
\end{equation}

Assuming [$^{12}$CO]/[$^{13}$CO]$\sim^{12}$C/$^{13}$C, the $^{12}$C/$^{13}$C ratio is also potentially an important factor controlling $R_{1213}$.
The $^{12}$C/$^{13}$C ratio is a measure of the ratio of primary to secondary nucleosynthetic components,
thus the ratio is expected to decrease as stellar populations evolve, and to depend on the star-formation history \citep[e.g.,][]{Audouze:1975jo}, the slope of IMF \citep[e.g.,][]{Romano:2017gk}, and the treatments of stellar rotation which affects mixing in stars \citep[e.g.,][]{Limongi:2018fr,Romano:2019ke}.
It is expected that, in the early evolutionary phase of stellar population, $^{12}$C/$^{13}$C is high due to efficient supply of $^{12}$C from core-collapse supernovae (SNe) of massive stars, and decrease due to $^{13}$C production from intermediate-mass asymptotic giant branch (AGB) stars, then increase due to $^{12}$C production from low-mass AGB stars, and decrease again due to type-Ia SNe \cite[e.g.,][]{Kobayashi:2011gg}.

There are theoretical and observational studies of $R_{1213}$ in normal galaxies, which contradict with scenarios (2)-(4) as a dominant factor changing $R_{1213}$.
\cite{Szucs:2014qe} self-consistently calculated $R_{1213}$ of isolated molecular clouds with hydrodynamical simulations taking into account chemical network, various interstellar radiation fields, and cooling and heating processes.
They concluded that the selective photodissociation has minimal effect on $R_{1213}$ and the chemical fractionation can cause a factor of $2-3$ decreases.
Other theoretical studies also predicted that the photodissociation of CO molecule is not selective since self-shielding is found to be not important compared to dust- and H$_2$-shielding \citep{Krumholz:2014oe,Safranek-Shrader:2017hm}.

The observed anti-correlations between $R_{1213}$ and $^{13}$CO/C$^{18}$O reported in various galaxies also disfavor the selective photodissociation scenario \citep[e.g.,][]{Paglione:2001if,Tan:2011dy}, because the rarer C$^{18}$O is expected to be more reduced than $^{13}$CO if the selective photodissociation is working.
It is reported that the radial gradient of $R_{1213}$ in the Milky Way is in the opposite sense of the $^{12}$C/$^{13}$C abundance gradient \citep{Paglione:2001if}.
\cite{Crocker:2012vs} compared several line ratios using six molecular lines and various galaxy properties of $^{12}$CO-rich ETGs, and found that correlations between $R_{1213}$ and the galaxy properties were similarly seen even when they used $^{12}$CO/HCN ratio (no difference in carbon isotope) instead.
This suggests that the chemical fractionation and $^{12}$C/$^{13}$C variation due to stellar nucleosynthesis are not the dominant processes generating the $R_{1213}$ variation in these ETGs.

However, the scenario (4) is likely to become important for extreme dusty SF galaxies such as local ULIRGs, starburst galaxies \citep{Sliwa:2017aq,Brown:2019ge}, and high-$z$ sub-milimetre galaxies (SMGs) \citep{Danielson:2013yl,Zhang:2018mc}.
In these galaxies, high $R_{1213}$ and extremely low $^{13}$CO/C$^{18}$O ($\sim1$) values have been found.
Given that $^{18}$O is produced in massive stars ($\gtrsim8$~M$_\odot$) and assuming $^{13}$CO/C$^{18}$O~$\approx$~$^{13}$C/$^{18}$O, the $^{13}$CO/C$^{18}$O is expected to be low for young systems or for top-heavy IMF.
Indeed, chemical evolution models of galaxies show that such extreme line ratio can be explained by top-heavy IMF \citep{Romano:2017gk,Zhang:2018mc,Romano:2019ke}.
The observed correlation between $R_{1213}$ and sSFR in this study may suggest that the stellar nucleosynthesis is also important since sSFR can be a measure of ratio of massive/young and less massive/old stellar populations.
The dominant process determining $R_{1213}$ may be different across the SF main sequence.
In order to investigate the effect of $^{12}$C/$^{13}$C on the $R_{1213}$ trend across the main sequence, further deep observations in optically-thin lines are required.

\subsection{Dependence of the six parameters on galaxy morphologies}
\label{sec:morph}

\begin{table*}[t]
    \caption{Median and 1st and 3rd quartiles of six key parameters in figure~\ref{fig:morph} of galaxy subsamples with different morphology\label{tab:morph}}
        \begin{center}
            \begin{tabular}{ccccccc}\hline
            RC3 morphology\footnotemark[$*$] & SA & SAB+SB & SAB & SB & w/ pec & w/o pec\\
            \# of galaxies & 27 & 33 & 20 & 13 & 10 & 53\\
            \hline
            $M_{\rm star}$ [$10^{10}$~M$_\odot$] & 3.8\footnotemark[$**$] & 3.8 & 3.9 & 3.1 & 2.8 & 4.1\\
             &  (1.8 -- 7.1)\footnotemark[$***$] &  (2.4 -- 7.3) &  (2.3 -- 6.3) &  (2.6 -- 11) &  (1.7 -- 6.9) &  (2.4 -- 7.3)\\
            SFR [M$_\odot$ yr$^{-1}$] & 1.9 & 3.1 & 2.7 & 4.7 & 2.0 & 2.4\\
             &  (1.2 -- 2.7) &  (1.8 -- 5.2) &  (1.8 -- 3.8) &  (2.0 -- 8.3) &  (1.3 -- 3.4) &  (1.4 -- 4.9)\\
            sSFR [$10^{-11}$~yr$^{-1}$] & 6.2 & 7.6 & 7.3 & 8.0 & 6.8 & 7.3\\
             &  (3.8 -- 8.3) &  (6.2 -- 9.9) &  (6.2 -- 8.9) &  (6.7 -- 16) &  (4.3 -- 12) &  (5.2 -- 8.8)\\
            \sigstai~[km/s] & 31 & 40 & 37 & 44 & 52 & 38\\
             &  (18 -- 47) &  (27 -- 53) &  (26 -- 45) &  (37 -- 58) &  (13 -- 59) &  (22 -- 46)\\
            $f_{60}/f_{100}$ & 0.41 & 0.47 & 0.43 & 0.48 & 0.49 & 0.43\\
             &  (0.37 -- 1.00) &  (0.40 -- 0.56) &  (0.39 -- 0.49) &  (0.46 -- 0.57) &  (0.41 -- 0.59) &  (0.39 -- 0.56)\\
            $R_{1213}$ & 12.2 & 15.3 & 16.9 & 14.1 & 17.1 & 14.1\\
             &  (10.5 -- 14.9) &  (13.2 -- 18.6) &  (14.2 -- 21.4) &  (12.9 -- 16.6) &  (13.2 -- 22.8) &  (10.8 -- 17.2)\\
            \hline
        \end{tabular}
    \end{center}
    \begin{tabnote}
    \footnotemark[$*$] There are three ``S'' galaxies among the 63 galaxies, i.e., no classification for bar structures.;
    \footnotemark[$**$] Median;
    \footnotemark[$***$] (1st quartile -- 3rd quartile)
    \end{tabnote}
\end{table*}

The turbulent and warm properties of interstellar medium (ISM) are a natural consequence for galaxies with high sSFR if star-formation feedbacks plays a major role in determining the ISM properties.
Theoretical studies showed that the molecular clouds are expanded due to feedback by \HII~regions and are destroyed or displaced by SNe explosion \citep[e.g.,][]{Baba:2017fk,Grisdale:2018qm}.
It is also shown that the observed scaling relations of molecular clouds are reproduced when these feedbacks are taken into account \citep[e.g.,][]{Grisdale:2018qm,Fujimoto:2019dt}.
However, both the spatial and velocity resolutions of our data are not high enough to resolve a single molecular cloud and to discriminate the effects of star-formation feedback from the other larger-scale processes.

The shorter dynamical (or free-fall) timescale than the star-formation timescale \citep[$1-3$~\%,][]{Kennicutt:1998fk,Leroy:2008fb} suggests that secular processes are important for star formation in galaxies.
It is widely known that the bar-structures are expected to drive gas inflow by removing angular momentum of the gas and consequently induce active star formation at the central regions of galaxies \citep[e.g.,][]{Wada:1995ly,Sakamoto:1999rt,Kuno:2007uq}.
Galaxy mergers are also a key process inducing gas condensation and triggering star formation in galaxies \citep[e.g.,][]{Sanders:1996ae,Hopkins:2005tv}.
While enhancement of star formation due to mergers or stellar bars may not be so significant \citep[$10-15$~\%,][]{Rodighiero:2011rv,Saintonge:2012nj}, both processes help to enhance turbulence and broaden the linewidth of CO emission in galaxies \citep{Sorai:2012ix,Maeda:2018gw,Sun:2018wz,Yajima:2019xz}.

\begin{figure*}[t]
\begin{center}
\includegraphics[width=150mm, bb=0 0 1395 723]{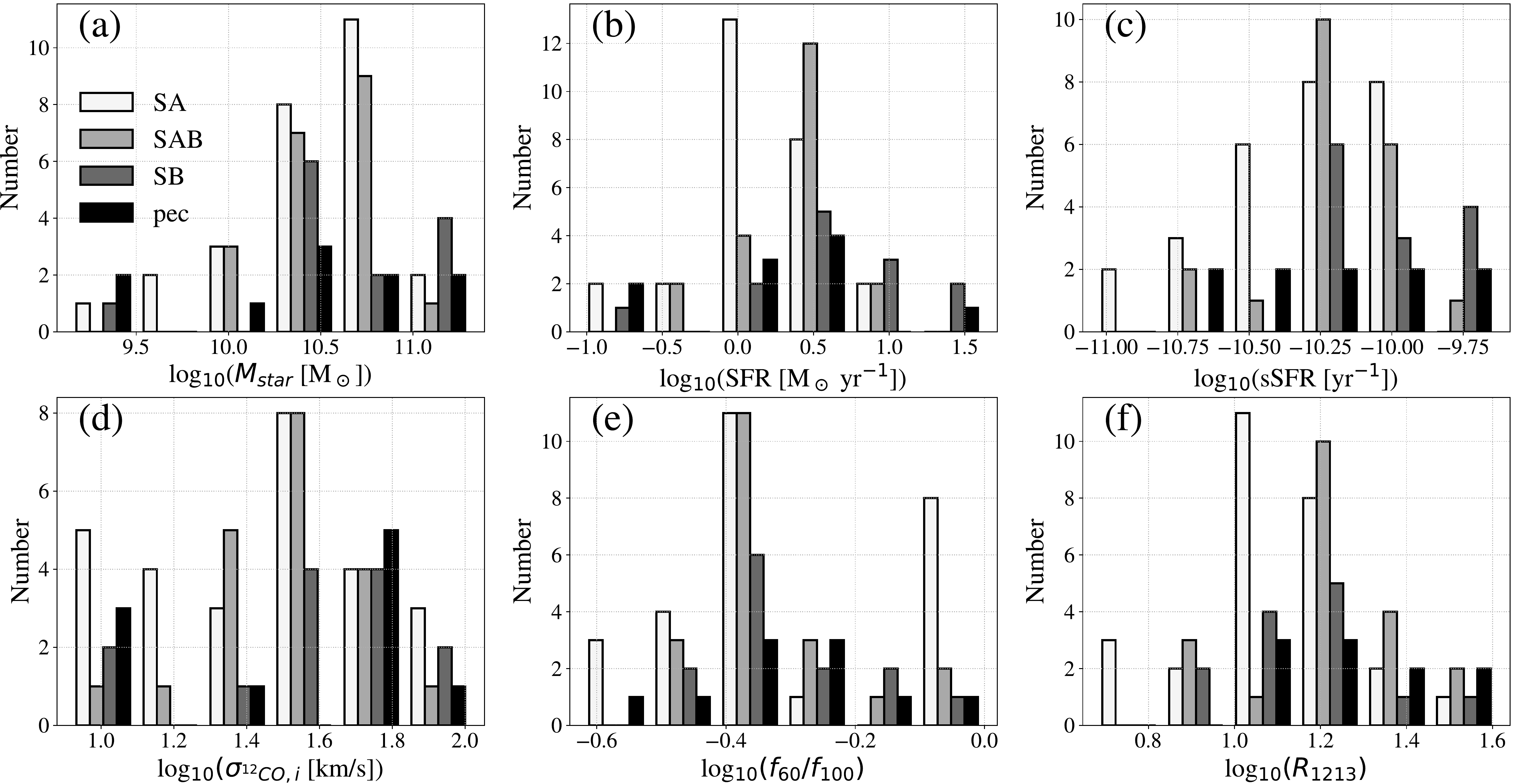}
\end{center}
\caption{
Histograms of the six key parameters for SA (white), SAB (light grey), SB (dark grey) galaxies and those with the ``pec'' suffix (black).
}
\label{fig:morph}
\end{figure*}

Figure~\ref{fig:morph} shows histograms of the six key values that are compared in figure~\ref{fig:comp} for subsamples with different types of bar structures \citep[SA/SAB/SB,][]{de-Vaucouleurs:1991pr} and table~\ref{tab:morph} summarizes medians and quartiles of these values for each subsample.
The galaxies categorized as ``pec'' in \cite{de-Vaucouleurs:1991pr} are also separately presented in the same figure and table.
According to the Kolmogorov-Smirnov (KS) test, there are no significant differences in $M_{\rm star}$ (p-value of $0.9$), sSFR ($0.07$), \sigstai~($0.3$), and $f_{60}/f_{100}$ ($0.3$) between barred (SAB+SB) and non-barred (SA) galaxies, whereas SFR and $R_{1213}$ tend to be higher for barred galaxies than non-barred galaxies, whose $p$-values are $4\times10^{-3}$ and $3\times10^{-2}$, respectively.
Although the $p$-values of the KS test are not small enough, the median values of sSFR, \sigstai, and $f_{60}/f_{100}$ are higher for barred galaxies than non-barred galaxies.
Therefore, the obtained tendency where galaxies with higher sSFR have broader \sigstai~in this study might be {\it partially} attributed to those barred galaxies.

In terms of the morphological peculiarities indicated with the suffix of ``pec'', on the other hand, we do not find a clear differences in any of the six values (figure~\ref{fig:morph} and table~\ref{tab:morph}).
Although no closely-interacting galaxies is included in our sample, the morphological peculiarities are generally indicative of galaxy interaction.
This suggests that the galaxy interaction is unlikely to be a major cause of the wider \sigstai~for galaxies with higher sSFR at least in our sample.
Therefore, high turbulence and temperature are likely to be primarily due to active star formation, and stellar bar might partially play some roles in driving the high turbulence of the molecular gas and the active star formation via efficient gas inflow.

\subsection{Implications of the variety of CO-to-H$_2$ conversion factor across the main sequence}
\label{sec:implication}

We adopted a constant $\alpha_{\rm CO}$ to estimate molecular gas mass of the COMING galaxies in this study, while the $R_{1213}$ variety suggests different $\alpha_{\rm CO}$ among the COMING galaxies.
\cite{Cormier:2018lt} showed that the relationship between $R_{1213}$ and $\alpha_{\rm CO}$ is different from galaxies to galaxies, while their non-LTE radiative transfer calculations with {\tt RADEX} \citep{van-der-Tak:2007qb} predicts a clear anti-correlation between these two values, which is ascribed to $\tau_{\rm ^{12}CO}$ variation \citep{Sandstrom:2013qp}.
Indeed, \cite{Accurso:2017il} found that $\alpha_{\rm CO}$ depends not only on metallicity but also on $\Delta$(MS).
In this subsection, we discuss a possible variety in $\alpha_{\rm CO}$ and its impact on our results.

Our data show that galaxies with high sSFR tend to have high $R_{1213}$, suggesting low $\tau_{\rm ^{12}CO}$ owing to high gas temperature and large linewidths.
According to magnetohydrodynamic simulations of molecular clouds, it is claimed that $\alpha_{\rm CO}$ scales with temperature and $^{12}$CO linewidth as $\propto T^{-1/2}$ and $\propto \Delta v^{-1/2}$, respectively \citep{Shetty:2011ai}.
The dust temperatures of the COMING galaxies inferred from $f_{60}/f_{100}$ and a conventional modified black-body model (opacity of $\kappa_0=0.009$~cm$^{2}$~g$^{-1}$ at $\nu_0$=200~GHz, and emissivity index of $\beta=1.8$) ranges $25-40$~K.
The \sigstai~ranges $\sim10-100$~km s$^{-1}$, although there also exists low-\sigstai~($<10$~km s$^{-1}$) galaxies in our sample.
Assuming gas temperature equals to dust temperature and $\Delta v \sim$~\sigstai, these varieties could impose an $\alpha_{\rm CO}$ variety by a factor of $\sim4$ {\it at most} for our sample.
Thus, the correlations between $R_{1213}$ and $\mu_{\rm mol}$ in figure~\ref{fig:basic}c would be actually weaker than the one obtained when using a constant $\alpha_{\rm CO}$.
With a sophisticated technique to derive $\alpha_{\rm CO}$ using CO, \HI~and dust maps of galaxies \citep{Leroy:2011rt,Sandstrom:2013qp}, $\alpha_{\rm CO}$ variations among the COMING galaxies are investigated in a forthcoming paper (Yasuda et al.~in prep.).

There are three caveats here:
(1) the resolutions of our data are not spatially or spectroscopically high enough to resolve a single molecular cloud,
(2) the ``luminosity-weighted'' dust temperature does not necessarily reflect the temperature of the bulk of molecular gas, and
(3) contribution from diffuse molecular gas.
The kpc-scale beam encompasses multiple molecular clouds and the observed velocity dispersion is likely to reflect inter-cloud velocity dispersion rather than that of a single cloud.
Additionally, the dust temperature derived by SED fitting is ``luminosity-weighted'' temperature rather than mass-weighted \citep{Scoville:2014rc,Scoville:2016de}.
But, the ranges of velocity dispersion and temperature we assumed here are not significantly different from the observed values for molecular clouds.
It is known that the velocity dispersions of molecular clouds including those at the central region of the Milky Way range $1-100$~km~s$^{-1}$ \citep[e.g.,][]{Oka:2001yv}.
The multi-transitions CO observations revealed that temperature of molecular clouds is reported to range from $10-20$~K for quiescent clouds to $15-100$~K for clouds with \HII~regions in M~33 \citep{Wilson:1997mb}.
The assumed varieties of $\Delta v$ and $T$ here are not so inadequate to consider the possible maximum variety in the $\alpha_{\rm CO}$ value.

For the third point, several studies of both the Milky-Way clouds and extragalactic objects suggest that non-negligible fraction of $^{12}$CO comes from diffuse molecular component \citep[$\sim30$~\%; e.g.,][]{Wilson:1994tn,Rosolowsky:2007is,Sawada:2012fv,Pety:2013rr,Morokuma-Matsui:2015cr,Maeda:2020gd,Liszt:2020jj}.
This diffuse molecular component is considered to consist of volume-filling diffuse molecular gas and/or envelope of molecular clouds \citep{Wilson:1994tn,Meier:2000wc,Pety:2013rr}.
Considering that the diffuse component tends to have large velocity dispersion \citep{Garcia-Burillo:1992as, Pety:2013rr, Caldu-Primo:2013lb}, the positive correlations of the sSFR-\sigstai and $R_{1213}$-\sigstai relations indicates a larger fraction of the diffuse component for galaxies with high sSFR and $R_{1213}$. 
The CO-to-H$_2$ conversion factor is reported to be similar in diffuse and dense molecular gas {\it on average} with a large scatter in the Milky Way because the small abundance ratio of [CO]/[H$_2$] in diffuse gas is compensated by a much higher brightness per CO molecule \citep{Liszt:2010te,Liszt:2012iz}.
However, it is important to investigate a fraction of the diffuse molecular-gas component of galaxies as a function of sSFR to understand the $\alpha_{\rm CO}$ variation across the main-sequence of SF galaxies.

\subsection{Implications for galaxy evolution on the SFR-Mass parameters space and high redshift galaxies}
\label{sec:galevo}

Our results suggest that there might be an $\alpha_{\rm CO}$ variety across the SF main sequence as a consequence of the differences in temperature and degree of turbulence of molecular gas in galaxies at least for the local Universe.
This further suggests that one may need to pay special attention when discussing either of which $\mu_{\rm mol}$ or SFE is the dominant factor controlling sSFR (or $\Delta$(MS)).
Previous studies found that $\Delta$(MS) and sSFR are correlated with both $\mu_{\rm mol}$ and SFE, i.e., galaxies with higher SFR at fixed $M_{\rm star}$ have more abundant molecular gas and are forming stars more efficiently from the molecular gas \citep{Saintonge:2012nj,Genzel:2015gn,Scoville:2017jw,Tacconi:2018pv}.
The possible $\alpha_{\rm CO}$ variation by a factor of $\sim4$ is unlikely to completely extinguish the reported correlations between sSFR and $\mu_{\rm mol}$ or SFE since the correlations are found over more than 2 orders of magnitude of the both values.
However, a magnitude relationship of the strength of the known positive correlations between sSFR and $\mu_{\rm mol}$ or SFE can be affected, since $M_{\rm mol}$ is numerator and denominator of $\mu_{\rm mol}$ and SFE, respectively.
If galaxies with higher sSFR tend to have lower $\alpha_{\rm CO}$, SFE would become relatively more important for the sSFR (or $\Delta$(MS)) variation than $\mu_{\rm mol}$, as claimed in \cite{Ellison:2020kf}.

The redshift evolution of $\alpha_{\rm CO}$ can be discussed in the same context combined with redshift evolution of metallicity.
sSFR of the main-sequence galaxies at fixed $M_{\rm star}$ is known to increase as redshift increases \citep[e.g.,][]{Whitaker:2012ok,Speagle:2014by}, making one expect higher turbulence and temperature of ISM in the higher-redshift galaxies from this study.
Indeed, high-redshift galaxies tend to have more highly turbulent molecular gas \citep{Genzel:2013pa,Ubler:2018ye,Girard:2019gz}, higher dust temperature \citep{Magdis:2012fh,Magnelli:2014zr,Schreiber:2018fe} and lower metallicity \citep[e.g.,][]{Savaglio:2005kq,Shapley:2005jk,Erb:2006jf,Liu:2008ry,Maiolino:2008ln,Mannucci:2009lp,Yuan:2013vg,Steidel:2014rw} than local galaxies.
With the Milky-Way type $\alpha_{\rm CO}$, a much larger $\mu_{\rm mol}$ value is reported in galaxies at $z\sim1-2$ \cite[$\sim50$~\%, e.g.,][and references therein]{Daddi:2010qq,Tacconi:2013qd,Genzel:2013pa,Riechers:2020nn,Tacconi:2020rs} than local galaxies \cite[$\lesssim5-10$~\%, e.g., ][]{Saintonge:2011hl,Saintonge:2017ve,Bolatto:2017jn,Sorai:2019hs}.
Since $\mu_{\rm mol}$ value is one of the key parameters to constraint on both star-formation and feedback models in cosmological simulations and semi-analytic models of galaxy formation \citep[e.g.,][]{Popping:2012gd,Popping:2014zm,Popping:2015fu,Morokuma-Matsui:2015vn,Saintonge:2017ve}, accurate $\alpha_{\rm CO}$ estimation at higher redshifts is essential.

Taking the three parameters (metallicity, velocity dispersion, and temperature) into consideration, we roughly estimate $\alpha_{\rm CO}$ of galaxies at high redshifts.
For galaxies with $M_{\rm star}\sim10^{10.5}$~M$_\odot$ at $z\sim1.5$, the metallicity, velocity dispersion of molecular gas, and dust temperature can be respectively $\sim0.2$-dex lower \citep{Yabe:2014dh,Zahid:2014om,Kashino:2017ar}\footnote{At the redshift, massive systems with $M_{\rm star}\sim10^{11}$~M$_\odot$ are known to be already enriched to the level observed in local counterparts \citep[][]{Zahid:2014om,Kashino:2017ar}.}, $\sim5$ times higher \citep{Genzel:2013pa}, and only $\sim1.2$~times higher than local counterparts.
In this case, the effects of metallicity and temperature are negligible \citep{Bolatto:2013vn} and $\alpha_{\rm CO}$ could become lower than the local value by a factor of 2.
Indeed, \cite{Carleton:2017jy} obtained comparable $\alpha_{\rm CO}$ for galaxies at $1<z<1.5$ to the local galaxies when they assume several models for redshift evolution of SFE.
Future high spatial- and velocity-resolution observations in $^{12}$CO of high redshift galaxies would shed light on the $\alpha_{\rm CO}$ variety across the SF main sequence of galaxies as a function of redshift.

\section{Summary}
\label{sec:summary}
Using the $^{12}$CO and $^{13}$CO mapping data obtained in the COMING project, we investigate the relationship between $R_{1213}$ and sSFR combined with other four parameters ($M_{\rm star}$, SFR, $f_{60}/f_{100}$, and \sigstai) of local star-forming galaxies to understand the molecular gas properties traced with $^{12}$CO line on the SFR-Mass relation of galaxies.
Owing to the stacking analysis, $^{13}$CO emission is successfully detected in 80 galaxies out of the 147 COMING galaxies.
The obtained results and implications are as follows:
\begin{itemize}
\item The error-weighted mean of $R_{1213}$ for the 80 COMING galaxies with measurements of $^{12}$CO and $^{13}$CO is $10.9$ with a standard deviation of $7.0$, which is consistent with previous studies on $R_{1213}$ of local SF galaxies (section~\ref{sec:results}, figure~\ref{fig:r1213_ssfr}).

\item $R_{1213}$ moderately correlates with sSFR with a Spearman's rank correlation coefficient of $0.46$ (section~\ref{sec:results}, figure~\ref{fig:r1213_ssfr}).

\item $R_{1213}$ variation is likely to be attributed to $\tau_{\rm ^{12}CO}$ variation, which is supported by the positive correlations of the $R_{1213}$-$f_{60}/f_{100}$ and $R_{1213}$-\sigstai~relations for galaxies with $R_{1213}$ measurements (section~\ref{sec:corrplots}, figures~\ref{fig:comp} and \ref{fig:corr}).

\item The correlations of the sSFR-$f_{60}/f_{100}$ and sSFR-\sigstai~relations for galaxies with $R_{1213}$ measurements suggest that high turbulence and temperature of ISM are related to active star formation (section~\ref{sec:corrplots}, figures~\ref{fig:comp} and \ref{fig:corr}).
Stellar bars might play some roles in enhancing the turbulence and in inducing the active star formation at the central regions of galaxies via gas inflow, which is supported by slightly larger median values of \sigstai~and sSFR for barred galaxies than non-barred galaxies in our sample although there is a large overlap between them (section~\ref{sec:morph}, figure~\ref{fig:morph} and table~\ref{tab:morph}).

\item The varieties in the dust temperature inferred from $f_{60}/f_{100}$ and the observed velocity dispersion of molecular gas in the COMING galaxies can impose $\alpha_{\rm CO}$ variation by a factor of $\sim4$ {\it at most} according to the relation of $\alpha_{\rm CO} \propto T^{-1/2} \Delta v^{-1/2}$ predicted in the numerical simulations of molecular clouds (sections~\ref{sec:causes} and \ref{sec:implication}).

\item The inferred $\alpha_{\rm CO}$ variation is not likely to completely extinguish the correlations between sSFR and $\mu_{\rm mol}$ or SFE reported in literatures.
However, a magnitude relationship of the strength of the positive correlations between sSFR and $\mu_{\rm mol}$ or SFE can be affected, since $M_{\rm mol}$ is numerator and denominator for $\mu_{\rm mol}$ and SFE, respectively (section~\ref{sec:galevo}).


\end{itemize}






\begin{ack}
We thank the anonymous referee for his/her comments which improve our paper.
This work was supported by JSPS KAKENHI Grant Numbers of 19J40004, 19H01931, 17H01110 and 19H05076.
This work has also been supported in part by the Sumitomo Foundation Fiscal 2018 Grant for Basic Science Research Projects (180923), and the Collaboration Funding of the Institute of Statistical Mathematics ``New Development of the Studies on Galaxy Evolution with a Method of Data Science''.
This publication makes use of data products from the Wide-field Infrared Survey Explorer, which is a joint project of the University of California, Los Angeles, and the Jet Propulsion Laboratory/California Institute of Technology, funded by the National Aeronautics and Space Administration.
\end{ack}

\bibliographystyle{apj}
\bibliography{/Users/kmatsui/Documents/Papers/myref_moro}

\begin{thebibliography}{}
\expandafter\ifx\csname natexlab\endcsname\relax\def\natexlab#1{#1}\fi

\bibitem[{{Aalto} {et~al.}(2010){Aalto}, {Beswick}, \&
  {J{\"u}tte}}]{Aalto:2010yx}
{Aalto}, S., {Beswick}, R., \& {J{\"u}tte}, E. 2010, \aap, 522, A59

\bibitem[{{Aalto} {et~al.}(1991){Aalto}, {Black}, {Johansson}, \&
  {Booth}}]{Aalto:1991bn}
{Aalto}, S., {Black}, J.~H., {Johansson}, L.~E.~B., \& {Booth}, R.~S. 1991,
  \aap, 249, 323

\bibitem[{{Aalto} {et~al.}(1995){Aalto}, {Booth}, {Black}, \&
  {Johansson}}]{Aalto:1995ir}
{Aalto}, S., {Booth}, R.~S., {Black}, J.~H., \& {Johansson}, L.~E.~B. 1995,
  \aap, 300, 369

\bibitem[{{Aalto} {et~al.}(1994){Aalto}, {Booth}, {Black}, {Koribalski}, \&
  {Wielebinski}}]{Aalto:1994yg}
{Aalto}, S., {Booth}, R.~S., {Black}, J.~H., {Koribalski}, B., \&
  {Wielebinski}, R. 1994, \aap, 286, 365

\bibitem[{{Aalto} {et~al.}(1997){Aalto}, {Radford}, {Scoville}, \&
  {Sargent}}]{Aalto:1997wk}
{Aalto}, S., {Radford}, S. J.~E., {Scoville}, N.~Z., \& {Sargent}, A.~I. 1997,
  \apjl, 475, L107

\bibitem[{{Accurso} {et~al.}(2017){Accurso}, {Saintonge}, {Catinella},
  {Cortese}, {Dav{\'e}}, {Dunsheath}, {Genzel}, {Gracia-Carpio}, {Heckman},
  {Jimmy}, {Kramer}, {Li}, {Lutz}, {Schiminovich}, {Schuster}, {Sternberg},
  {Sturm}, {Tacconi}, {Tran}, \& {Wang}}]{Accurso:2017il}
{Accurso}, G., {Saintonge}, A., {Catinella}, B., {et~al.} 2017, \mnras, 470,
  4750

\bibitem[{{Alatalo} {et~al.}(2015){Alatalo}, {Crocker}, {Aalto}, {Davis},
  {Nyland}, {Bureau}, {Duc}, {Krajnovi{\'c}}, \& {Young}}]{Alatalo:2015ca}
{Alatalo}, K., {Crocker}, A.~F., {Aalto}, S., {et~al.} 2015, \mnras, 450, 3874

\bibitem[{{Audouze} {et~al.}(1975){Audouze}, {Lequeux}, \&
  {Vigroux}}]{Audouze:1975jo}
{Audouze}, J., {Lequeux}, J., \& {Vigroux}, L. 1975, \aap, 43, 71

\bibitem[{{Baba} {et~al.}(2017){Baba}, {Morokuma-Matsui}, \&
  {Saitoh}}]{Baba:2017fk}
{Baba}, J., {Morokuma-Matsui}, K., \& {Saitoh}, T.~R. 2017, \mnras, 464, 246

\bibitem[{{Becker} \& {Freudling}(1991)}]{Becker:1991da}
{Becker}, R., \& {Freudling}, W. 1991, \aap, 251, 454

\bibitem[{{B{\'e}thermin} {et~al.}(2018){B{\'e}thermin}, {Greve}, {De Breuck},
  {Vieira}, {Aravena}, {Chapman}, {Chen}, {Dong}, {Hayward}, {Hezaveh},
  {Marrone}, {Narayanan}, {Phadke}, {Reuter}, {Spilker}, {Stark}, {Strandet},
  \& {Wei{\ss}}}]{Bethermin:2018jq}
{B{\'e}thermin}, M., {Greve}, T.~R., {De Breuck}, C., {et~al.} 2018, \aap, 620,
  A115

\bibitem[{{Blitz} {et~al.}(1984){Blitz}, {Magnani}, \& {Mundy}}]{Blitz:1984ky}
{Blitz}, L., {Magnani}, L., \& {Mundy}, L. 1984, \apjl, 282, L9

\bibitem[{{Bolatto} {et~al.}(2013){Bolatto}, {Wolfire}, \&
  {Leroy}}]{Bolatto:2013vn}
{Bolatto}, A.~D., {Wolfire}, M., \& {Leroy}, A.~K. 2013, \araa, 51, 207

\bibitem[{{Bolatto} {et~al.}(2017){Bolatto}, {Wong}, {Utomo}, {Blitz}, {Vogel},
  {S{\'a}nchez}, {Barrera-Ballesteros}, {Cao}, {Colombo}, {Dannerbauer},
  {Garc{\'\i}a-Benito}, {Herrera-Camus}, {Husemann}, {Kalinova}, {Leroy},
  {Leung}, {Levy}, {Mast}, {Ostriker}, {Rosolowsky}, {Sandstrom}, {Teuben},
  {van de Ven}, \& {Walter}}]{Bolatto:2017jn}
{Bolatto}, A.~D., {Wong}, T., {Utomo}, D., {et~al.} 2017, \apj, 846, 159

\bibitem[{{Braine} {et~al.}(1993){Braine}, {Combes}, \& {van
  Driel}}]{Braine:1993iw}
{Braine}, J., {Combes}, F., \& {van Driel}, W. 1993, \aap, 280, 451

\bibitem[{{Brown} \& {Wilson}(2019)}]{Brown:2019ge}
{Brown}, T., \& {Wilson}, C.~D. 2019, \apj, 879, 17

\bibitem[{{Cald{\'u}-Primo} {et~al.}(2013){Cald{\'u}-Primo}, {Schruba},
  {Walter}, {Leroy}, {Sandstrom}, {de Blok}, {Ianjamasimanana}, \&
  {Mogotsi}}]{Caldu-Primo:2013lb}
{Cald{\'u}-Primo}, A., {Schruba}, A., {Walter}, F., {et~al.} 2013, \aj, 146,
  150

\bibitem[{{Cao} {et~al.}(2017){Cao}, {Wong}, {Xue}, {Bolatto}, {Blitz},
  {Vogel}, {Leroy}, \& {Rosolowsky}}]{Cao:2017bf}
{Cao}, Y., {Wong}, T., {Xue}, R., {et~al.} 2017, \apj, 847, 33

\bibitem[{{Carleton} {et~al.}(2017){Carleton}, {Cooper}, {Bolatto}, {Bournaud},
  {Combes}, {Freundlich}, {Garcia-Burillo}, {Genzel}, {Neri}, {Tacconi}, {Sand
  strom}, {Weiner}, \& {Weiss}}]{Carleton:2017jy}
{Carleton}, T., {Cooper}, M.~C., {Bolatto}, A.~D., {et~al.} 2017, \mnras, 467,
  4886

\bibitem[{{Casasola} {et~al.}(2017){Casasola}, {Cassar{\`a}}, {Bianchi},
  {Verstocken}, {Xilouris}, {Magrini}, {Smith}, {De Looze}, {Galametz},
  {Madden}, {Baes}, {Clark}, {Davies}, {De Vis}, {Evans}, {Fritz}, {Galliano},
  {Jones}, {Mosenkov}, {Viaene}, \& {Ysard}}]{Casasola:2017zv}
{Casasola}, V., {Cassar{\`a}}, L.~P., {Bianchi}, S., {et~al.} 2017, \aap, 605,
  A18

\bibitem[{{Casoli} {et~al.}(1992){Casoli}, {Dupraz}, \&
  {Combes}}]{Casoli:1992zy}
{Casoli}, F., {Dupraz}, C., \& {Combes}, F. 1992, \aap, 264, 55

\bibitem[{{Cormier} {et~al.}(2018){Cormier}, {Bigiel}, {Jim{\'e}nez-Donaire},
  {Leroy}, {Gallagher}, {Usero}, {Sandstrom}, {Bolatto}, {Hughes}, {Kramer},
  {Krumholz}, {Meier}, {Murphy}, {Pety}, {Rosolowsky}, {Schinnerer}, {Schruba},
  {Sliwa}, \& {Walter}}]{Cormier:2018lt}
{Cormier}, D., {Bigiel}, F., {Jim{\'e}nez-Donaire}, M.~J., {et~al.} 2018,
  \mnras, 475, 3909

\bibitem[{{Crocker} {et~al.}(2012){Crocker}, {Krips}, {Bureau}, {Young},
  {Davis}, {Bayet}, {Alatalo}, {Blitz}, {Bois}, {Bournaud}, {Cappellari},
  {Davies}, {de Zeeuw}, {Duc}, {Emsellem}, {Khochfar}, {Krajnovi{\'c}},
  {Kuntschner}, {Lablanche}, {McDermid}, {Morganti}, {Naab}, {Oosterloo},
  {Sarzi}, {Scott}, {Serra}, \& {Weijmans}}]{Crocker:2012vs}
{Crocker}, A., {Krips}, M., {Bureau}, M., {et~al.} 2012, \mnras, 421, 1298

\bibitem[{{Daddi} {et~al.}(2007){Daddi}, {Dickinson}, {Morrison}, {Chary},
  {Cimatti}, {Elbaz}, {Frayer}, {Renzini}, {Pope}, {Alexander}, {Bauer},
  {Giavalisco}, {Huynh}, {Kurk}, \& {Mignoli}}]{Daddi:2007un}
{Daddi}, E., {Dickinson}, M., {Morrison}, G., {et~al.} 2007, \apj, 670, 156

\bibitem[{{Daddi} {et~al.}(2010){Daddi}, {Bournaud}, {Walter}, {Dannerbauer},
  {Carilli}, {Dickinson}, {Elbaz}, {Morrison}, {Riechers}, {Onodera}, {Salmi},
  {Krips}, \& {Stern}}]{Daddi:2010qq}
{Daddi}, E., {Bournaud}, F., {Walter}, F., {et~al.} 2010, \apj, 713, 686

\bibitem[{{Danielson} {et~al.}(2013){Danielson}, {Swinbank}, {Smail}, {Bayet},
  {van der Werf}, {Cox}, {Edge}, {Henkel}, \& {Ivison}}]{Danielson:2013yl}
{Danielson}, A.~L.~R., {Swinbank}, A.~M., {Smail}, I., {et~al.} 2013, \mnras,
  436, 2793

\bibitem[{{Davis}(2014)}]{Davis:2014tq}
{Davis}, T.~A. 2014, \mnras, 445, 2378

\bibitem[{{de Vaucouleurs} {et~al.}(1991){de Vaucouleurs}, {de Vaucouleurs},
  {Corwin}, {Buta}, {Paturel}, \& {Fouqu{\'e}}}]{de-Vaucouleurs:1991pr}
{de Vaucouleurs}, G., {de Vaucouleurs}, A., {Corwin}, Jr., H.~G., {et~al.}
  1991, {Third Reference Catalogue of Bright Galaxies. Volume I: Explanations
  and references. Volume II: Data for galaxies between 0$^{h}$ and 12$^{h}$.
  Volume III: Data for galaxies between 12$^{h}$ and 24$^{h}$.}

\bibitem[{{Dickman}(1978)}]{Dickman:1978nn}
{Dickman}, R.~L. 1978, \apjs, 37, 407

\bibitem[{{Eckart} {et~al.}(1990){Eckart}, {Cameron}, {Rothermel}, {Wild},
  {Zinnecker}, {Rydbeck}, {Olberg}, \& {Wiklind}}]{Eckart:1990rp}
{Eckart}, A., {Cameron}, M., {Rothermel}, H., {et~al.} 1990, \apj, 363, 451

\bibitem[{{Elbaz} {et~al.}(2007){Elbaz}, {Daddi}, {Le Borgne}, {Dickinson},
  {Alexander}, {Chary}, {Starck}, {Brand t}, {Kitzbichler}, {MacDonald},
  {Nonino}, {Popesso}, {Stern}, \& {Vanzella}}]{Elbaz:2007ub}
{Elbaz}, D., {Daddi}, E., {Le Borgne}, D., {et~al.} 2007, \aap, 468, 33

\bibitem[{{Ellison} {et~al.}(2020{\natexlab{a}}){Ellison}, {Thorp}, {Pan},
  {Lin}, {Scudder}, {Bluck}, {S{\'a}nchez}, \& {Sargent}}]{Ellison:2020dm}
{Ellison}, S.~L., {Thorp}, M.~D., {Pan}, H.-A., {et~al.} 2020{\natexlab{a}},
  \mnras, 492, 6027

\bibitem[{{Ellison} {et~al.}(2020{\natexlab{b}}){Ellison}, {Thorp}, {Lin},
  {Pan}, {Bluck}, {Scudder}, {Teimoorinia}, {S{\'a}nchez}, \&
  {Sargent}}]{Ellison:2020kf}
{Ellison}, S.~L., {Thorp}, M.~D., {Lin}, L., {et~al.} 2020{\natexlab{b}},
  \mnras, 493, L39

\bibitem[{{Encrenaz} {et~al.}(1979){Encrenaz}, {Stark}, {Combes}, \&
  {Wilson}}]{Encrenaz:1979dw}
{Encrenaz}, P.~J., {Stark}, A.~A., {Combes}, F., \& {Wilson}, R.~W. 1979, \aap,
  78, L1

\bibitem[{{Erb} {et~al.}(2006){Erb}, {Shapley}, {Pettini}, {Steidel}, {Reddy},
  \& {Adelberger}}]{Erb:2006jf}
{Erb}, D.~K., {Shapley}, A.~E., {Pettini}, M., {et~al.} 2006, \apj, 644, 813

\bibitem[{{Frerking} {et~al.}(1982){Frerking}, {Langer}, \&
  {Wilson}}]{Frerking:1982zq}
{Frerking}, M.~A., {Langer}, W.~D., \& {Wilson}, R.~W. 1982, \apj, 262, 590

\bibitem[{{Fujimoto} {et~al.}(2019){Fujimoto}, {Chevance}, {Haydon},
  {Krumholz}, \& {Kruijssen}}]{Fujimoto:2019dt}
{Fujimoto}, Y., {Chevance}, M., {Haydon}, D.~T., {Krumholz}, M.~R., \&
  {Kruijssen}, J.~M.~D. 2019, \mnras, 487, 1717

\bibitem[{{Garay} {et~al.}(1993){Garay}, {Mardones}, \&
  {Mirabel}}]{Garay:1993wo}
{Garay}, G., {Mardones}, D., \& {Mirabel}, I.~F. 1993, \aap, 277, 405

\bibitem[{{Garcia-Burillo} {et~al.}(1992){Garcia-Burillo}, {Guelin},
  {Cernicharo}, \& {Dahlem}}]{Garcia-Burillo:1992as}
{Garcia-Burillo}, S., {Guelin}, M., {Cernicharo}, J., \& {Dahlem}, M. 1992,
  \aap, 266, 21

\bibitem[{{Genzel} {et~al.}(2013){Genzel}, {Tacconi}, {Kurk}, {Wuyts},
  {Combes}, {Freundlich}, {Bolatto}, {Cooper}, {Neri}, {Nordon}, {Bournaud},
  {Burkert}, {Comerford}, {Cox}, {Davis}, {F{\"o}rster Schreiber},
  {Garc{\'\i}a-Burillo}, {Gracia-Carpio}, {Lutz}, {Naab}, {Newman},
  {Saintonge}, {Shapiro Griffin}, {Shapley}, {Sternberg}, \&
  {Weiner}}]{Genzel:2013pa}
{Genzel}, R., {Tacconi}, L.~J., {Kurk}, J., {et~al.} 2013, \apj, 773, 68

\bibitem[{{Genzel} {et~al.}(2015){Genzel}, {Tacconi}, {Lutz}, {Saintonge},
  {Berta}, {Magnelli}, {Combes}, {Garc{\'\i}a-Burillo}, {Neri}, {Bolatto},
  {Contini}, {Lilly}, {Boissier}, {Boone}, {Bouch{\'e}}, {Bournaud}, {Burkert},
  {Carollo}, {Colina}, {Cooper}, {Cox}, {Feruglio}, {F{\"o}rster Schreiber},
  {Freundlich}, {Gracia-Carpio}, {Juneau}, {Kovac}, {Lippa}, {Naab}, {Salome},
  {Renzini}, {Sternberg}, {Walter}, {Weiner}, {Weiss}, \&
  {Wuyts}}]{Genzel:2015gn}
{Genzel}, R., {Tacconi}, L.~J., {Lutz}, D., {et~al.} 2015, \apj, 800, 20

\bibitem[{{Girard} {et~al.}(2019){Girard}, {Dessauges-Zavadsky}, {Combes},
  {Chisholm}, {Patr{\'\i}cio}, {Richard}, \& {Schaerer}}]{Girard:2019gz}
{Girard}, M., {Dessauges-Zavadsky}, M., {Combes}, F., {et~al.} 2019, \aap, 631,
  A91

\bibitem[{{Gordon} \& {Burton}(1976)}]{Gordon:1976ly}
{Gordon}, M.~A., \& {Burton}, W.~B. 1976, \apj, 208, 346

\bibitem[{{Grisdale} {et~al.}(2018){Grisdale}, {Agertz}, {Renaud}, \&
  {Romeo}}]{Grisdale:2018qm}
{Grisdale}, K., {Agertz}, O., {Renaud}, F., \& {Romeo}, A.~B. 2018, \mnras,
  479, 3167

\bibitem[{{Halfen} {et~al.}(2017){Halfen}, {Woolf}, \&
  {Ziurys}}]{Halfen:2017sd}
{Halfen}, D.~T., {Woolf}, N.~J., \& {Ziurys}, L.~M. 2017, \apj, 845, 158

\bibitem[{{Henkel} {et~al.}(2010){Henkel}, {Downes}, {Wei{\ss}}, {Riechers}, \&
  {Walter}}]{Henkel:2010av}
{Henkel}, C., {Downes}, D., {Wei{\ss}}, A., {Riechers}, D., \& {Walter}, F.
  2010, \aap, 516, A111

\bibitem[{{Herrero-Illana} {et~al.}(2019){Herrero-Illana}, {Privon}, {Evans},
  {D{\'\i}az-Santos}, {P{\'e}rez-Torres}, {U}, {Alberdi}, {Iwasawa}, {Armus},
  {Aalto}, {Mazzarella}, {Chu}, {Sanders}, {Barcos-Mu{\~n}oz}, {Charmandaris},
  {Linden}, {Yoon}, {Frayer}, {Inami}, {Kim}, {Borish}, {Conway}, {Murphy},
  {Song}, {Stierwalt}, \& {Surace}}]{Herrero-Illana:2019lc}
{Herrero-Illana}, R., {Privon}, G.~C., {Evans}, A.~S., {et~al.} 2019, \aap,
  628, A71

\bibitem[{{Hirota} {et~al.}(2010){Hirota}, {Kuno}, {Sato}, {Nakanishi},
  {Tosaki}, \& {Sorai}}]{Hirota:2010af}
{Hirota}, A., {Kuno}, N., {Sato}, N., {et~al.} 2010, \pasj, 62, 1261

\bibitem[{{Hopkins} {et~al.}(2005){Hopkins}, {Hernquist}, {Cox}, {Di Matteo},
  {Martini}, {Robertson}, \& {Springel}}]{Hopkins:2005tv}
{Hopkins}, P.~F., {Hernquist}, L., {Cox}, T.~J., {et~al.} 2005, \apj, 630, 705

\bibitem[{{Kashino} {et~al.}(2017){Kashino}, {Silverman}, {Sanders},
  {Kartaltepe}, {Daddi}, {Renzini}, {Valentino}, {Rodighiero}, {Juneau},
  {Kewley}, {Zahid}, {Arimoto}, {Nagao}, {Chu}, {Sugiyama}, {Civano}, {Ilbert},
  {Kajisawa}, {Le F{\`e}vre}, {Maier}, {Masters}, {Miyaji}, {Onodera},
  {Puglisi}, \& {Taniguchi}}]{Kashino:2017ar}
{Kashino}, D., {Silverman}, J.~D., {Sanders}, D., {et~al.} 2017, \apj, 835, 88

\bibitem[{{Kennicutt}(1998)}]{Kennicutt:1998fk}
{Kennicutt}, Jr., R.~C. 1998, \apj, 498, 541

\bibitem[{{Kikumoto} {et~al.}(1998){Kikumoto}, {Taniguchi}, {Nakai},
  {Ishizuki}, {Matsushita}, \& {Kawabe}}]{Kikumoto:1998xw}
{Kikumoto}, T., {Taniguchi}, Y., {Nakai}, N., {et~al.} 1998, \pasj, 50, 309

\bibitem[{{Knapp} \& {Bowers}(1988)}]{Knapp:1988gf}
{Knapp}, G.~R., \& {Bowers}, P.~F. 1988, \apj, 331, 974

\bibitem[{{Kobayashi} {et~al.}(2011){Kobayashi}, {Karakas}, \&
  {Umeda}}]{Kobayashi:2011gg}
{Kobayashi}, C., {Karakas}, A.~I., \& {Umeda}, H. 2011, \mnras, 414, 3231

\bibitem[{{Koyama} {et~al.}(2017){Koyama}, {Koyama}, {Yamashita},
  {Morokuma-Matsui}, {Matsuhara}, {Nakagawa}, {Hayashi}, {Kodama}, {Shimakawa},
  {Suzuki}, {Tadaki}, {Tanaka}, \& {Yamamoto}}]{Koyama:2017lf}
{Koyama}, S., {Koyama}, Y., {Yamashita}, T., {et~al.} 2017, \apj, 847, 137

\bibitem[{{Koyama} {et~al.}(2019){Koyama}, {Koyama}, {Yamashita}, {Hayashi},
  {Matsuhara}, {Nakagawa}, {Namiki}, {Suzuki}, {Fukagawa}, {Kodama}, {Lin},
  {Morokuma-Matsui}, {Shimakawa}, \& {Tanaka}}]{Koyama:2019vy}
---. 2019, \apj, 874, 142

\bibitem[{{Kroupa}(2001)}]{Kroupa:2001xf}
{Kroupa}, P. 2001, \mnras, 322, 231

\bibitem[{{Kroupa} \& {Weidner}(2003)}]{Kroupa:2003do}
{Kroupa}, P., \& {Weidner}, C. 2003, \apj, 598, 1076

\bibitem[{{Krumholz}(2014)}]{Krumholz:2014oe}
{Krumholz}, M.~R. 2014, \mnras, 437, 1662

\bibitem[{{Kuno} {et~al.}(2007){Kuno}, {Sato}, {Nakanishi}, {Hirota}, {Tosaki},
  {Shioya}, {Sorai}, {Nakai}, {Nishiyama}, \& {Vila-Vilar{\'o}}}]{Kuno:2007uq}
{Kuno}, N., {Sato}, N., {Nakanishi}, H., {et~al.} 2007, \pasj, 59, 117

\bibitem[{{Lada} {et~al.}(1994){Lada}, {Lada}, {Clemens}, \&
  {Bally}}]{Lada:1994ya}
{Lada}, C.~J., {Lada}, E.~A., {Clemens}, D.~P., \& {Bally}, J. 1994, \apj, 429,
  694

\bibitem[{{Lee} \& {Chung}(2018)}]{Lee:2018rt}
{Lee}, B., \& {Chung}, A. 2018, \apjl, 866, L10

\bibitem[{{Leitherer} {et~al.}(1999){Leitherer}, {Schaerer}, {Goldader},
  {Delgado}, {Robert}, {Kune}, {de Mello}, {Devost}, \&
  {Heckman}}]{Leitherer:1999ye}
{Leitherer}, C., {Schaerer}, D., {Goldader}, J.~D., {et~al.} 1999, \apjs, 123,
  3

\bibitem[{{Leroy} {et~al.}(2008){Leroy}, {Walter}, {Brinks}, {Bigiel}, {de
  Blok}, {Madore}, \& {Thornley}}]{Leroy:2008fb}
{Leroy}, A.~K., {Walter}, F., {Brinks}, E., {et~al.} 2008, \aj, 136, 2782

\bibitem[{{Leroy} {et~al.}(2011){Leroy}, {Bolatto}, {Gordon}, {Sand strom},
  {Gratier}, {Rosolowsky}, {Engelbracht}, {Mizuno}, {Corbelli}, {Fukui}, \&
  {Kawamura}}]{Leroy:2011rt}
{Leroy}, A.~K., {Bolatto}, A., {Gordon}, K., {et~al.} 2011, \apj, 737, 12

\bibitem[{{Limongi} \& {Chieffi}(2018)}]{Limongi:2018fr}
{Limongi}, M., \& {Chieffi}, A. 2018, \apjs, 237, 13

\bibitem[{{Lin} {et~al.}(2017){Lin}, {Belfiore}, {Pan}, {Bothwell}, {Hsieh},
  {Huang}, {Xiao}, {S{\'a}nchez}, {Hsieh}, {Masters}, {Ramya}, {Lin}, {Hsu},
  {Li}, {Maiolino}, {Bundy}, {Bizyaev}, {Drory}, {Ibarra-Medel}, {Lacerna},
  {Haines}, {Smethurst}, {Stark}, \& {Thomas}}]{Lin:2017ee}
{Lin}, L., {Belfiore}, F., {Pan}, H.-A., {et~al.} 2017, \apj, 851, 18

\bibitem[{{Liszt}(2020)}]{Liszt:2020jj}
{Liszt}, H.~S. 2020, arXiv e-prints, arXiv:2005.10270

\bibitem[{{Liszt} \& {Pety}(2012)}]{Liszt:2012iz}
{Liszt}, H.~S., \& {Pety}, J. 2012, \aap, 541, A58

\bibitem[{{Liszt} {et~al.}(2010){Liszt}, {Pety}, \& {Lucas}}]{Liszt:2010te}
{Liszt}, H.~S., {Pety}, J., \& {Lucas}, R. 2010, \aap, 518, A45

\bibitem[{{Liu} {et~al.}(2008){Liu}, {Shapley}, {Coil}, {Brinchmann}, \&
  {Ma}}]{Liu:2008ry}
{Liu}, X., {Shapley}, A.~E., {Coil}, A.~L., {Brinchmann}, J., \& {Ma}, C.-P.
  2008, \apj, 678, 758

\bibitem[{{Maeda} {et~al.}(2018){Maeda}, {Ohta}, {Fujimoto}, {Habe}, \&
  {Baba}}]{Maeda:2018gw}
{Maeda}, F., {Ohta}, K., {Fujimoto}, Y., {Habe}, A., \& {Baba}, J. 2018, \pasj,
  70, 37

\bibitem[{{Maeda} {et~al.}(2020){Maeda}, {Ohta}, {Fujimoto}, {Habe}, \&
  {Ushio}}]{Maeda:2020gd}
{Maeda}, F., {Ohta}, K., {Fujimoto}, Y., {Habe}, A., \& {Ushio}, K. 2020,
  \mnras, arXiv:2005.03019

\bibitem[{{Magdis} {et~al.}(2012){Magdis}, {Daddi}, {B{\'e}thermin}, {Sargent},
  {Elbaz}, {Pannella}, {Dickinson}, {Dannerbauer}, {da Cunha}, {Walter},
  {Rigopoulou}, {Charmandaris}, {Hwang}, \& {Kartaltepe}}]{Magdis:2012fh}
{Magdis}, G.~E., {Daddi}, E., {B{\'e}thermin}, M., {et~al.} 2012, \apj, 760, 6

\bibitem[{{Magnelli} {et~al.}(2014){Magnelli}, {Lutz}, {Saintonge}, {Berta},
  {Santini}, {Symeonidis}, {Altieri}, {Andreani}, {Aussel}, {B{\'e}thermin},
  {Bock}, {Bongiovanni}, {Cepa}, {Cimatti}, {Conley}, {Daddi}, {Elbaz},
  {F{\"o}rster Schreiber}, {Genzel}, {Ivison}, {Le Floc'h}, {Magdis},
  {Maiolino}, {Nordon}, {Oliver}, {Page}, {P{\'e}rez Garc{\'\i}a}, {Poglitsch},
  {Popesso}, {Pozzi}, {Riguccini}, {Rodighiero}, {Rosario}, {Roseboom},
  {Sanchez-Portal}, {Scott}, {Sturm}, {Tacconi}, {Valtchanov}, {Wang}, \&
  {Wuyts}}]{Magnelli:2014zr}
{Magnelli}, B., {Lutz}, D., {Saintonge}, A., {et~al.} 2014, \aap, 561, A86

\bibitem[{{Maiolino} {et~al.}(2008){Maiolino}, {Nagao}, {Grazian}, {Cocchia},
  {Marconi}, {Mannucci}, {Cimatti}, {Pipino}, {Ballero}, {Calura}, {Chiappini},
  {Fontana}, {Granato}, {Matteucci}, {Pastorini}, {Pentericci}, {Risaliti},
  {Salvati}, \& {Silva}}]{Maiolino:2008ln}
{Maiolino}, R., {Nagao}, T., {Grazian}, A., {et~al.} 2008, \aap, 488, 463

\bibitem[{{Mannucci} {et~al.}(2009){Mannucci}, {Cresci}, {Maiolino}, {Marconi},
  {Pastorini}, {Pozzetti}, {Gnerucci}, {Risaliti}, {Schneider}, {Lehnert}, \&
  {Salvati}}]{Mannucci:2009lp}
{Mannucci}, F., {Cresci}, G., {Maiolino}, R., {et~al.} 2009, \mnras, 398, 1915

\bibitem[{{Martin} {et~al.}(2005){Martin}, {Fanson}, {Schiminovich},
  {Morrissey}, {Friedman}, {Barlow}, {Conrow}, {Grange}, {Jelinsky},
  {Milliard}, {Siegmund}, {Bianchi}, {Byun}, {Donas}, {Forster}, {Heckman},
  {Lee}, {Madore}, {Malina}, {Neff}, {Rich}, {Small}, {Surber}, {Szalay},
  {Welsh}, \& {Wyder}}]{Martin:2005wd}
{Martin}, D.~C., {Fanson}, J., {Schiminovich}, D., {et~al.} 2005, \apjl, 619,
  L1

\bibitem[{{Matsuki} {et~al.}(2017){Matsuki}, {Koyama}, {Nakagawa}, \&
  {Takita}}]{Matsuki:2017nl}
{Matsuki}, Y., {Koyama}, Y., {Nakagawa}, T., \& {Takita}, S. 2017, \mnras, 466,
  2517

\bibitem[{{Matsushita} {et~al.}(1998){Matsushita}, {Kohno}, {Vila-Vilaro},
  {Tosaki}, \& {Kawabe}}]{Matsushita:1998si}
{Matsushita}, S., {Kohno}, K., {Vila-Vilaro}, B., {Tosaki}, T., \& {Kawabe}, R.
  1998, \apj, 495, 267

\bibitem[{{Meier} \& {Turner}(2004)}]{Meier:2004cy}
{Meier}, D.~S., \& {Turner}, J.~L. 2004, \aj, 127, 2069

\bibitem[{{Meier} {et~al.}(2000){Meier}, {Turner}, \& {Hurt}}]{Meier:2000wc}
{Meier}, D.~S., {Turner}, J.~L., \& {Hurt}, R.~L. 2000, \apj, 531, 200

\bibitem[{{Milam} {et~al.}(2005){Milam}, {Savage}, {Brewster}, {Ziurys}, \&
  {Wyckoff}}]{Milam:2005yb}
{Milam}, S.~N., {Savage}, C., {Brewster}, M.~A., {Ziurys}, L.~M., \& {Wyckoff},
  S. 2005, \apj, 634, 1126

\bibitem[{{Morokuma-Matsui} \& {Baba}(2015)}]{Morokuma-Matsui:2015vn}
{Morokuma-Matsui}, K., \& {Baba}, J. 2015, \mnras, 454, 3792

\bibitem[{{Morokuma-Matsui} {et~al.}(2015){Morokuma-Matsui}, {Sorai},
  {Watanabe}, \& {Kuno}}]{Morokuma-Matsui:2015cr}
{Morokuma-Matsui}, K., {Sorai}, K., {Watanabe}, Y., \& {Kuno}, N. 2015, \pasj,
  67, 2

\bibitem[{{Noeske} {et~al.}(2007){Noeske}, {Weiner}, {Faber}, {Papovich},
  {Koo}, {Somerville}, {Bundy}, {Conselice}, {Newman}, {Schiminovich}, {Le
  Floc'h}, {Coil}, {Rieke}, {Lotz}, {Primack}, {Barmby}, {Cooper}, {Davis},
  {Ellis}, {Fazio}, {Guhathakurta}, {Huang}, {Kassin}, {Martin}, {Phillips},
  {Rich}, {Small}, {Willmer}, \& {Wilson}}]{Noeske:2007lr}
{Noeske}, K.~G., {Weiner}, B.~J., {Faber}, S.~M., {et~al.} 2007, \apjl, 660,
  L43

\bibitem[{{Oka} {et~al.}(2001){Oka}, {Hasegawa}, {Sato}, {Tsuboi}, {Miyazaki},
  \& {Sugimoto}}]{Oka:2001yv}
{Oka}, T., {Hasegawa}, T., {Sato}, F., {et~al.} 2001, \apj, 562, 348

\bibitem[{{Paglione} {et~al.}(2001){Paglione}, {Wall}, {Young}, {Heyer},
  {Richard}, {Goldstein}, {Kaufman}, {Nantais}, \& {Perry}}]{Paglione:2001if}
{Paglione}, T. A.~D., {Wall}, W.~F., {Young}, J.~S., {et~al.} 2001, \apjs, 135,
  183

\bibitem[{{Papadopoulos} \& {Seaquist}(1998)}]{Papadopoulos:1998fh}
{Papadopoulos}, P.~P., \& {Seaquist}, E.~R. 1998, \apj, 492, 521

\bibitem[{{Papadopoulos} \& {Seaquist}(1999)}]{Papadopoulos:1999wa}
---. 1999, \apj, 516, 114

\bibitem[{{Pety} {et~al.}(2013){Pety}, {Schinnerer}, {Leroy}, {Hughes},
  {Meidt}, {Colombo}, {Dumas}, {Garc{\'{\i}}a-Burillo}, {Schuster}, {Kramer},
  {Dobbs}, \& {Thompson}}]{Pety:2013rr}
{Pety}, J., {Schinnerer}, E., {Leroy}, A.~K., {et~al.} 2013, \apj, 779, 43

\bibitem[{{Polk} {et~al.}(1988){Polk}, {Knapp}, {Stark}, \&
  {Wilson}}]{Polk:1988ly}
{Polk}, K.~S., {Knapp}, G.~R., {Stark}, A.~A., \& {Wilson}, R.~W. 1988, \apj,
  332, 432

\bibitem[{{Popping} {et~al.}(2015){Popping}, {Behroozi}, \&
  {Peeples}}]{Popping:2015fu}
{Popping}, G., {Behroozi}, P.~S., \& {Peeples}, M.~S. 2015, \mnras, 449, 477

\bibitem[{{Popping} {et~al.}(2012){Popping}, {Caputi}, {Somerville}, \&
  {Trager}}]{Popping:2012gd}
{Popping}, G., {Caputi}, K.~I., {Somerville}, R.~S., \& {Trager}, S.~C. 2012,
  \mnras, 425, 2386

\bibitem[{{Popping} {et~al.}(2014){Popping}, {Somerville}, \&
  {Trager}}]{Popping:2014zm}
{Popping}, G., {Somerville}, R.~S., \& {Trager}, S.~C. 2014, \mnras, 442, 2398

\bibitem[{{Rickard} {et~al.}(1977){Rickard}, {Palmer}, {Morris}, {Turner}, \&
  {Zuckerman}}]{Rickard:1977ox}
{Rickard}, L.~J., {Palmer}, P., {Morris}, M., {Turner}, B.~E., \& {Zuckerman},
  B. 1977, \apj, 213, 673

\bibitem[{{Rickard} {et~al.}(1975){Rickard}, {Palmer}, {Morris}, {Zuckerman},
  \& {Turner}}]{Rickard:1975if}
{Rickard}, L.~J., {Palmer}, P., {Morris}, M., {Zuckerman}, B., \& {Turner},
  B.~E. 1975, \apjl, 199, L75

\bibitem[{{Riechers} {et~al.}(2020){Riechers}, {Boogaard}, {Decarli},
  {Gonzalez-Lopez}, {Smail}, {Walter}, {Aravena}, {Carilli}, {Cortes}, {Cox},
  {Diaz-Santos}, {Hodge}, {Inami}, {Ivison}, {Kaasinen}, {Wagg}, {Weiss}, \&
  {van der Werf}}]{Riechers:2020nn}
{Riechers}, D.~A., {Boogaard}, L.~A., {Decarli}, R., {et~al.} 2020, arXiv
  e-prints, arXiv:2005.09653

\bibitem[{{Rodighiero} {et~al.}(2011){Rodighiero}, {Daddi}, {Baronchelli},
  {Cimatti}, {Renzini}, {Aussel}, {Popesso}, {Lutz}, {Andreani}, {Berta},
  {Cava}, {Elbaz}, {Feltre}, {Fontana}, {F{\"o}rster Schreiber},
  {Franceschini}, {Genzel}, {Grazian}, {Gruppioni}, {Ilbert}, {Le Floch},
  {Magdis}, {Magliocchetti}, {Magnelli}, {Maiolino}, {McCracken}, {Nordon},
  {Poglitsch}, {Santini}, {Pozzi}, {Riguccini}, {Tacconi}, {Wuyts}, \&
  {Zamorani}}]{Rodighiero:2011rv}
{Rodighiero}, G., {Daddi}, E., {Baronchelli}, I., {et~al.} 2011, \apjl, 739,
  L40

\bibitem[{{R{\"o}llig} \& {Ossenkopf}(2013)}]{Rollig:2013xw}
{R{\"o}llig}, M., \& {Ossenkopf}, V. 2013, \aap, 550, A56

\bibitem[{{Romano} {et~al.}(2019){Romano}, {Matteucci}, {Zhang}, {Ivison}, \&
  {Ventura}}]{Romano:2019ke}
{Romano}, D., {Matteucci}, F., {Zhang}, Z.-Y., {Ivison}, R.~J., \& {Ventura},
  P. 2019, \mnras, 490, 2838

\bibitem[{{Romano} {et~al.}(2017){Romano}, {Matteucci}, {Zhang},
  {Papadopoulos}, \& {Ivison}}]{Romano:2017gk}
{Romano}, D., {Matteucci}, F., {Zhang}, Z.~Y., {Papadopoulos}, P.~P., \&
  {Ivison}, R.~J. 2017, \mnras, 470, 401

\bibitem[{{Rosolowsky} {et~al.}(2007){Rosolowsky}, {Keto}, {Matsushita}, \&
  {Willner}}]{Rosolowsky:2007is}
{Rosolowsky}, E., {Keto}, E., {Matsushita}, S., \& {Willner}, S.~P. 2007, \apj,
  661, 830

\bibitem[{{Safranek-Shrader} {et~al.}(2017){Safranek-Shrader}, {Krumholz},
  {Kim}, {Ostriker}, {Klein}, {Li}, {McKee}, \&
  {Stone}}]{Safranek-Shrader:2017hm}
{Safranek-Shrader}, C., {Krumholz}, M.~R., {Kim}, C.-G., {et~al.} 2017, \mnras,
  465, 885

\bibitem[{{Sage}(1990)}]{Sage:1990yz}
{Sage}, L.~J. 1990, \aap, 239, 125

\bibitem[{{Sage} \& {Isbell}(1991)}]{Sage:1991rj}
{Sage}, L.~J., \& {Isbell}, D.~W. 1991, \aap, 247, 320

\bibitem[{{Saintonge} {et~al.}(2011){Saintonge}, {Kauffmann}, {Kramer},
  {Tacconi}, {Buchbender}, {Catinella}, {Fabello}, {Graci{\'a}-Carpio}, {Wang},
  {Cortese}, {Fu}, {Genzel}, {Giovanelli}, {Guo}, {Haynes}, {Heckman},
  {Krumholz}, {Lemonias}, {Li}, {Moran}, {Rodriguez-Fernandez}, {Schiminovich},
  {Schuster}, \& {Sievers}}]{Saintonge:2011hl}
{Saintonge}, A., {Kauffmann}, G., {Kramer}, C., {et~al.} 2011, \mnras, 415, 32

\bibitem[{{Saintonge} {et~al.}(2012){Saintonge}, {Tacconi}, {Fabello}, {Wang},
  {Catinella}, {Genzel}, {Graci{\'a}-Carpio}, {Kramer}, {Moran}, {Heckman},
  {Schiminovich}, {Schuster}, \& {Wuyts}}]{Saintonge:2012nj}
{Saintonge}, A., {Tacconi}, L.~J., {Fabello}, S., {et~al.} 2012, \apj, 758, 73

\bibitem[{{Saintonge} {et~al.}(2017){Saintonge}, {Catinella}, {Tacconi},
  {Kauffmann}, {Genzel}, {Cortese}, {Dav{\'e}}, {Fletcher},
  {Graci{\'a}-Carpio}, {Kramer}, {Heckman}, {Janowiecki}, {Lutz}, {Rosario},
  {Schiminovich}, {Schuster}, {Wang}, {Wuyts}, {Borthakur}, {Lamperti}, \&
  {Roberts-Borsani}}]{Saintonge:2017ve}
{Saintonge}, A., {Catinella}, B., {Tacconi}, L.~J., {et~al.} 2017, \apjs, 233,
  22

\bibitem[{{Sakamoto} {et~al.}(1999){Sakamoto}, {Okumura}, {Ishizuki}, \&
  {Scoville}}]{Sakamoto:1999rt}
{Sakamoto}, K., {Okumura}, S.~K., {Ishizuki}, S., \& {Scoville}, N.~Z. 1999,
  \apj, 525, 691

\bibitem[{{Sakamoto} {et~al.}(1997){Sakamoto}, {Handa}, {Sofue}, {Honma}, \&
  {Sorai}}]{Sakamoto:1997gr}
{Sakamoto}, S., {Handa}, T., {Sofue}, Y., {Honma}, M., \& {Sorai}, K. 1997,
  \apj, 475, 134

\bibitem[{{Sakamoto} {et~al.}(1994){Sakamoto}, {Hayashi}, {Hasegawa}, {Handa},
  \& {Oka}}]{Sakamoto:1994di}
{Sakamoto}, S., {Hayashi}, M., {Hasegawa}, T., {Handa}, T., \& {Oka}, T. 1994,
  \apj, 425, 641

\bibitem[{{Salpeter}(1955)}]{Salpeter:1955bz}
{Salpeter}, E.~E. 1955, \apj, 121, 161

\bibitem[{{Sanders} {et~al.}(2003){Sanders}, {Mazzarella}, {Kim}, {Surace}, \&
  {Soifer}}]{Sanders:2003ci}
{Sanders}, D.~B., {Mazzarella}, J.~M., {Kim}, D.~C., {Surace}, J.~A., \&
  {Soifer}, B.~T. 2003, \aj, 126, 1607

\bibitem[{{Sanders} \& {Mirabel}(1996)}]{Sanders:1996ae}
{Sanders}, D.~B., \& {Mirabel}, I.~F. 1996, \araa, 34, 749

\bibitem[{{Sandqvist} {et~al.}(1988){Sandqvist}, {Elfhag}, \&
  {Jorsater}}]{Sandqvist:1988fh}
{Sandqvist}, A., {Elfhag}, T., \& {Jorsater}, S. 1988, \aap, 201, 223

\bibitem[{{Sandstrom} {et~al.}(2013){Sandstrom}, {Leroy}, {Walter}, {Bolatto},
  {Croxall}, {Draine}, {Wilson}, {Wolfire}, {Calzetti}, {Kennicutt}, {Aniano},
  {Donovan Meyer}, {Usero}, {Bigiel}, {Brinks}, {de Blok}, {Crocker}, {Dale},
  {Engelbracht}, {Galametz}, {Groves}, {Hunt}, {Koda}, {Kreckel}, {Linz},
  {Meidt}, {Pellegrini}, {Rix}, {Roussel}, {Schinnerer}, {Schruba}, {Schuster},
  {Skibba}, {van der Laan}, {Appleton}, {Armus}, {Brandl}, {Gordon}, {Hinz},
  {Krause}, {Montiel}, {Sauvage}, {Schmiedeke}, {Smith}, \&
  {Vigroux}}]{Sandstrom:2013qp}
{Sandstrom}, K.~M., {Leroy}, A.~K., {Walter}, F., {et~al.} 2013, \apj, 777, 5

\bibitem[{{Savaglio} {et~al.}(2005){Savaglio}, {Glazebrook}, {Le Borgne},
  {Juneau}, {Abraham}, {Chen}, {Crampton}, {McCarthy}, {Carlberg}, {Marzke},
  {Roth}, {J{\o}rgensen}, \& {Murowinski}}]{Savaglio:2005kq}
{Savaglio}, S., {Glazebrook}, K., {Le Borgne}, D., {et~al.} 2005, \apj, 635,
  260

\bibitem[{{Sawada} {et~al.}(2012){Sawada}, {Hasegawa}, {Sugimoto}, {Koda}, \&
  {Handa}}]{Sawada:2012fv}
{Sawada}, T., {Hasegawa}, T., {Sugimoto}, M., {Koda}, J., \& {Handa}, T. 2012,
  \apj, 752, 118

\bibitem[{{Schreiber} {et~al.}(2018){Schreiber}, {Elbaz}, {Pannella}, {Ciesla},
  {Wang}, \& {Franco}}]{Schreiber:2018fe}
{Schreiber}, C., {Elbaz}, D., {Pannella}, M., {et~al.} 2018, \aap, 609, A30

\bibitem[{{Schruba} {et~al.}(2011){Schruba}, {Leroy}, {Walter}, {Bigiel},
  {Brinks}, {de Blok}, {Dumas}, {Kramer}, {Rosolowsky}, {Sandstrom},
  {Schuster}, {Usero}, {Weiss}, \& {Wiesemeyer}}]{Schruba:2011zr}
{Schruba}, A., {Leroy}, A.~K., {Walter}, F., {et~al.} 2011, \aj, 142, 37

\bibitem[{{Scoville} {et~al.}(2014){Scoville}, {Aussel}, {Sheth}, {Scott},
  {Sanders}, {Ivison}, {Pope}, {Capak}, {Vanden Bout}, {Manohar}, {Kartaltepe},
  {Robertson}, \& {Lilly}}]{Scoville:2014rc}
{Scoville}, N., {Aussel}, H., {Sheth}, K., {et~al.} 2014, \apj, 783, 84

\bibitem[{{Scoville} {et~al.}(2016){Scoville}, {Sheth}, {Aussel}, {Vanden
  Bout}, {Capak}, {Bongiorno}, {Casey}, {Murchikova}, {Koda},
  {{\'A}lvarez-M{\'a}rquez}, {Lee}, {Laigle}, {McCracken}, {Ilbert}, {Pope},
  {Sanders}, {Chu}, {Toft}, {Ivison}, \& {Manohar}}]{Scoville:2016de}
{Scoville}, N., {Sheth}, K., {Aussel}, H., {et~al.} 2016, \apj, 820, 83

\bibitem[{{Scoville} {et~al.}(2017){Scoville}, {Lee}, {Vanden Bout},
  {Diaz-Santos}, {Sanders}, {Darvish}, {Bongiorno}, {Casey}, {Murchikova},
  {Koda}, {Capak}, {Vlahakis}, {Ilbert}, {Sheth}, {Morokuma-Matsui}, {Ivison},
  {Aussel}, {Laigle}, {McCracken}, {Armus}, {Pope}, {Toft}, \&
  {Masters}}]{Scoville:2017jw}
{Scoville}, N., {Lee}, N., {Vanden Bout}, P., {et~al.} 2017, \apj, 837, 150

\bibitem[{{Shapley} {et~al.}(2005){Shapley}, {Coil}, {Ma}, \&
  {Bundy}}]{Shapley:2005jk}
{Shapley}, A.~E., {Coil}, A.~L., {Ma}, C.-P., \& {Bundy}, K. 2005, \apj, 635,
  1006

\bibitem[{{Sheth} {et~al.}(2010){Sheth}, {Regan}, {Hinz}, {Gil de Paz},
  {Men{\'e}ndez-Delmestre}, {Mu{\~n}oz-Mateos}, {Seibert}, {Kim},
  {Laurikainen}, {Salo}, {Gadotti}, {Laine}, {Mizusawa}, {Armus},
  {Athanassoula}, {Bosma}, {Buta}, {Capak}, {Jarrett}, {Elmegreen},
  {Elmegreen}, {Knapen}, {Koda}, {Helou}, {Ho}, {Madore}, {Masters},
  {Mobasher}, {Ogle}, {Peng}, {Schinnerer}, {Surace}, {Zaritsky},
  {Comer{\'o}n}, {de Swardt}, {Meidt}, {Kasliwal}, \& {Aravena}}]{Sheth:2010mz}
{Sheth}, K., {Regan}, M., {Hinz}, J.~L., {et~al.} 2010, \pasp, 122, 1397

\bibitem[{{Shetty} {et~al.}(2011){Shetty}, {Glover}, {Dullemond}, {Ostriker},
  {Harris}, \& {Klessen}}]{Shetty:2011ai}
{Shetty}, R., {Glover}, S.~C., {Dullemond}, C.~P., {et~al.} 2011, \mnras, 415,
  3253

\bibitem[{{Sliwa} {et~al.}(2017){Sliwa}, {Wilson}, {Aalto}, \&
  {Privon}}]{Sliwa:2017aq}
{Sliwa}, K., {Wilson}, C.~D., {Aalto}, S., \& {Privon}, G.~C. 2017, \apjl, 840,
  L11

\bibitem[{{Solomon} \& {de Zafra}(1975)}]{Solomon:1975jr}
{Solomon}, P.~M., \& {de Zafra}, R. 1975, \apjl, 199, L79

\bibitem[{{Solomon} {et~al.}(1979){Solomon}, {Scoville}, \&
  {Sanders}}]{Solomon:1979zd}
{Solomon}, P.~M., {Scoville}, N.~Z., \& {Sanders}, D.~B. 1979, \apjl, 232, L89

\bibitem[{{Sorai} {et~al.}(2012){Sorai}, {Kuno}, {Nishiyama}, {Watanabe},
  {Matsui}, {Habe}, {Hirota}, {Ishihara}, \& {Nakai}}]{Sorai:2012ix}
{Sorai}, K., {Kuno}, N., {Nishiyama}, K., {et~al.} 2012, \pasj, 64, 51

\bibitem[{{Sorai} {et~al.}(2019){Sorai}, {Kuno}, {Muraoka}, {Miyamoto},
  {Kaneko}, {Nakanishi}, {Nakai}, {Yanagitani}, {Tanaka}, {Sato}, {Salak},
  {Umei}, {Morokuma-Matsui}, {Matsumoto}, {Ueno}, {Pan}, {Noma}, {Takeuchi},
  {Yoda}, {Kuroda}, {Yasuda}, {Yajima}, {Oi}, {Shibata}, {Seta}, {Watanabe},
  {Kita}, {Komatsuzaki}, {Kajikawa}, {Yashima}, {Cooray}, {Baji}, {Segawa},
  {Tashiro}, {Takeda}, {Kishida}, {Hatakeyama}, {Tomiyasu}, \&
  {Saita}}]{Sorai:2019hs}
{Sorai}, K., {Kuno}, N., {Muraoka}, K., {et~al.} 2019, \pasj, 71, S14

\bibitem[{{Speagle} {et~al.}(2014){Speagle}, {Steinhardt}, {Capak}, \&
  {Silverman}}]{Speagle:2014by}
{Speagle}, J.~S., {Steinhardt}, C.~L., {Capak}, P.~L., \& {Silverman}, J.~D.
  2014, \apjs, 214, 15

\bibitem[{{Spilker} {et~al.}(2014){Spilker}, {Marrone}, {Aguirre}, {Aravena},
  {Ashby}, {B{\'e}thermin}, {Bradford}, {Bothwell}, {Brodwin}, {Carlstrom},
  {Chapman}, {Crawford}, {de Breuck}, {Fassnacht}, {Gonzalez}, {Greve},
  {Gullberg}, {Hezaveh}, {Holzapfel}, {Husband}, {Ma}, {Malkan}, {Murphy},
  {Reichardt}, {Rotermund}, {Stalder}, {Stark}, {Strandet}, {Vieira},
  {Wei{\ss}}, \& {Welikala}}]{Spilker:2014lc}
{Spilker}, J.~S., {Marrone}, D.~P., {Aguirre}, J.~E., {et~al.} 2014, \apj, 785,
  149

\bibitem[{{Stark} \& {Carlson}(1984)}]{Stark:1984av}
{Stark}, A.~A., \& {Carlson}, E.~R. 1984, \apj, 279, 122

\bibitem[{{Steidel} {et~al.}(2014){Steidel}, {Rudie}, {Strom}, {Pettini},
  {Reddy}, {Shapley}, {Trainor}, {Erb}, {Turner}, {Konidaris}, {Kulas}, {Mace},
  {Matthews}, \& {McLean}}]{Steidel:2014rw}
{Steidel}, C.~C., {Rudie}, G.~C., {Strom}, A.~L., {et~al.} 2014, \apj, 795, 165

\bibitem[{{Sun} {et~al.}(2018){Sun}, {Leroy}, {Schruba}, {Rosolowsky},
  {Hughes}, {Kruijssen}, {Meidt}, {Schinnerer}, {Blanc}, {Bigiel}, {Bolatto},
  {Chevance}, {Groves}, {Herrera}, {Hygate}, {Pety}, {Querejeta}, {Usero}, \&
  {Utomo}}]{Sun:2018wz}
{Sun}, J., {Leroy}, A.~K., {Schruba}, A., {et~al.} 2018, \apj, 860, 172

\bibitem[{{Sz{\H{u}}cs} {et~al.}(2014){Sz{\H{u}}cs}, {Glover}, \&
  {Klessen}}]{Szucs:2014qe}
{Sz{\H{u}}cs}, L., {Glover}, S. C.~O., \& {Klessen}, R.~S. 2014, \mnras, 445,
  4055

\bibitem[{{Tacconi} {et~al.}(2020){Tacconi}, {Genzel}, \&
  {Sternberg}}]{Tacconi:2020rs}
{Tacconi}, L.~J., {Genzel}, R., \& {Sternberg}, A. 2020, arXiv e-prints,
  arXiv:2003.06245

\bibitem[{{Tacconi} {et~al.}(2013){Tacconi}, {Neri}, {Genzel}, {Combes},
  {Bolatto}, {Cooper}, {Wuyts}, {Bournaud}, {Burkert}, {Comerford}, {Cox},
  {Davis}, {F{\"o}rster Schreiber}, {Garc{\'{\i}}a-Burillo}, {Gracia-Carpio},
  {Lutz}, {Naab}, {Newman}, {Omont}, {Saintonge}, {Shapiro Griffin}, {Shapley},
  {Sternberg}, \& {Weiner}}]{Tacconi:2013qd}
{Tacconi}, L.~J., {Neri}, R., {Genzel}, R., {et~al.} 2013, \apj, 768, 74

\bibitem[{{Tacconi} {et~al.}(2018){Tacconi}, {Genzel}, {Saintonge}, {Combes},
  {Garc{\'\i}a-Burillo}, {Neri}, {Bolatto}, {Contini}, {F{\"o}rster Schreiber},
  {Lilly}, {Lutz}, {Wuyts}, {Accurso}, {Boissier}, {Boone}, {Bouch{\'e}},
  {Bournaud}, {Burkert}, {Carollo}, {Cooper}, {Cox}, {Feruglio}, {Freundlich},
  {Herrera-Camus}, {Juneau}, {Lippa}, {Naab}, {Renzini}, {Salome}, {Sternberg},
  {Tadaki}, {{\"U}bler}, {Walter}, {Weiner}, \& {Weiss}}]{Tacconi:2018pv}
{Tacconi}, L.~J., {Genzel}, R., {Saintonge}, A., {et~al.} 2018, \apj, 853, 179

\bibitem[{{Tan} {et~al.}(2011){Tan}, {Gao}, {Zhang}, \& {Xia}}]{Tan:2011dy}
{Tan}, Q.-H., {Gao}, Y., {Zhang}, Z.-Y., \& {Xia}, X.-Y. 2011, Research in
  Astronomy and Astrophysics, 11, 787

\bibitem[{{Taniguchi} \& {Ohyama}(1998)}]{Taniguchi:1998yi}
{Taniguchi}, Y., \& {Ohyama}, Y. 1998, \apjl, 507, L121

\bibitem[{{{\"U}bler} {et~al.}(2018){{\"U}bler}, {Genzel}, {Tacconi},
  {F{\"o}rster Schreiber}, {Neri}, {Contursi}, {Belli}, {Nelson}, {Lang},
  {Shimizu}, {Davies}, {Herrera-Camus}, {Lutz}, {Plewa}, {Price}, {Schuster},
  {Sternberg}, {Tadaki}, {Wisnioski}, \& {Wuyts}}]{Ubler:2018ye}
{{\"U}bler}, H., {Genzel}, R., {Tacconi}, L.~J., {et~al.} 2018, \apjl, 854, L24

\bibitem[{{van der Tak} {et~al.}(2007){van der Tak}, {Black}, {Sch{\"o}ier},
  {Jansen}, \& {van Dishoeck}}]{van-der-Tak:2007qb}
{van der Tak}, F.~F.~S., {Black}, J.~H., {Sch{\"o}ier}, F.~L., {Jansen}, D.~J.,
  \& {van Dishoeck}, E.~F. 2007, \aap, 468, 627

\bibitem[{{Vila-Vilaro} {et~al.}(2015){Vila-Vilaro}, {Cepa}, \&
  {Zabludoff}}]{Vila-Vilaro:2015ip}
{Vila-Vilaro}, B., {Cepa}, J., \& {Zabludoff}, A. 2015, \apjs, 218, 28

\bibitem[{{Wada} \& {Habe}(1995)}]{Wada:1995ly}
{Wada}, K., \& {Habe}, A. 1995, \mnras, 277, 433

\bibitem[{{Watanabe} {et~al.}(2011){Watanabe}, {Sorai}, {Kuno}, \&
  {Habe}}]{Watanabe:2011na}
{Watanabe}, Y., {Sorai}, K., {Kuno}, N., \& {Habe}, A. 2011, \mnras, 411, 1409

\bibitem[{{Weliachew} {et~al.}(1988){Weliachew}, {Casoli}, \&
  {Combes}}]{Weliachew:1988tx}
{Weliachew}, L., {Casoli}, F., \& {Combes}, F. 1988, \aap, 199, 29

\bibitem[{{Wen} {et~al.}(2013){Wen}, {Wu}, {Zhu}, {Lam}, {Wu}, {Wicker}, \&
  {Zhao}}]{Wen:2013wt}
{Wen}, X.-Q., {Wu}, H., {Zhu}, Y.-N., {et~al.} 2013, \mnras, 433, 2946

\bibitem[{{Whitaker} {et~al.}(2012){Whitaker}, {van Dokkum}, {Brammer}, \&
  {Franx}}]{Whitaker:2012ok}
{Whitaker}, K.~E., {van Dokkum}, P.~G., {Brammer}, G., \& {Franx}, M. 2012,
  \apjl, 754, L29

\bibitem[{{Wild} {et~al.}(1997){Wild}, {Eckart}, \& {Wiklind}}]{Wild:1997dm}
{Wild}, W., {Eckart}, A., \& {Wiklind}, T. 1997, \aap, 322, 419

\bibitem[{{Wilson} \& {Walker}(1994)}]{Wilson:1994tn}
{Wilson}, C.~D., \& {Walker}, C.~E. 1994, \apj, 432, 148

\bibitem[{{Wilson} {et~al.}(1997){Wilson}, {Walker}, \&
  {Thornley}}]{Wilson:1997mb}
{Wilson}, C.~D., {Walker}, C.~E., \& {Thornley}, M.~D. 1997, \apj, 483, 210

\bibitem[{{Wright} {et~al.}(2010){Wright}, {Eisenhardt}, {Mainzer}, {Ressler},
  {Cutri}, {Jarrett}, {Kirkpatrick}, {Padgett}, {McMillan}, {Skrutskie},
  {Stanford}, {Cohen}, {Walker}, {Mather}, {Leisawitz}, {Gautier}, {McLean},
  {Benford}, {Lonsdale}, {Blain}, {Mendez}, {Irace}, {Duval}, {Liu}, {Royer},
  {Heinrichsen}, {Howard}, {Shannon}, {Kendall}, {Walsh}, {Larsen}, {Cardon},
  {Schick}, {Schwalm}, {Abid}, {Fabinsky}, {Naes}, \& {Tsai}}]{Wright:2010oi}
{Wright}, E.~L., {Eisenhardt}, P. R.~M., {Mainzer}, A.~K., {et~al.} 2010, \aj,
  140, 1868

\bibitem[{{Wright} {et~al.}(1993){Wright}, {Ishizuki}, {Turner}, {Ho}, \&
  {Lo}}]{Wright:1993ic}
{Wright}, M.~C.~H., {Ishizuki}, S., {Turner}, J.~L., {Ho}, P.~T.~P., \& {Lo},
  K.~Y. 1993, \apj, 406, 470

\bibitem[{{Wuyts} {et~al.}(2011){Wuyts}, {F{\"o}rster Schreiber}, {Lutz},
  {Nordon}, {Berta}, {Altieri}, {Andreani}, {Aussel}, {Bongiovanni}, {Cepa},
  {Cimatti}, {Daddi}, {Elbaz}, {Genzel}, {Koekemoer}, {Magnelli}, {Maiolino},
  {McGrath}, {P{\'e}rez Garc{\'{\i}}a}, {Poglitsch}, {Popesso}, {Pozzi},
  {Sanchez-Portal}, {Sturm}, {Tacconi}, \& {Valtchanov}}]{Wuyts:2011uq}
{Wuyts}, S., {F{\"o}rster Schreiber}, N.~M., {Lutz}, D., {et~al.} 2011, \apj,
  738, 106

\bibitem[{{Xie} {et~al.}(1994){Xie}, {Young}, \& {Schloerb}}]{Xie:1994do}
{Xie}, S., {Young}, J., \& {Schloerb}, F.~P. 1994, \apj, 421, 434

\bibitem[{{Yabe} {et~al.}(2014){Yabe}, {Ohta}, {Iwamuro}, {Akiyama}, {Tamura},
  {Yuma}, {Kimura}, {Takato}, {Moritani}, {Sumiyoshi}, {Maihara}, {Silverman},
  {Dalton}, {Lewis}, {Bonfield}, {Lee}, {Curtis-Lake}, {Macaulay}, \&
  {Clarke}}]{Yabe:2014dh}
{Yabe}, K., {Ohta}, K., {Iwamuro}, F., {et~al.} 2014, \mnras, 437, 3647

\bibitem[{{Yajima} {et~al.}(2019){Yajima}, {Sorai}, {Kuno}, {Muraoka},
  {Miyamoto}, {Kaneko}, {Nakanishi}, {Nakai}, {Tanaka}, {Sato}, {Salak},
  {Morokuma-Matsui}, {Matsumoto}, {Pan}, {Noma}, {Takeuchi}, {Yoda}, {Kuroda},
  {Yasuda}, {Oi}, {Shibata}, {Seta}, {Watanabe}, {Kita}, {Komatsuzaki},
  {Kajikawa}, \& {Yashima}}]{Yajima:2019xz}
{Yajima}, Y., {Sorai}, K., {Kuno}, N., {et~al.} 2019, \pasj, 71, S13

\bibitem[{{Young} \& {Sanders}(1986)}]{Young:1986hh}
{Young}, J.~S., \& {Sanders}, D.~B. 1986, \apj, 302, 680

\bibitem[{{Young} \& {Scoville}(1982)}]{Young:1982sd}
{Young}, J.~S., \& {Scoville}, N. 1982, \apj, 258, 467

\bibitem[{{Young} \& {Scoville}(1984)}]{Young:1984sl}
{Young}, J.~S., \& {Scoville}, N.~Z. 1984, \apj, 287, 153

\bibitem[{{Yuan} {et~al.}(2013){Yuan}, {Kewley}, \& {Richard}}]{Yuan:2013vg}
{Yuan}, T.~T., {Kewley}, L.~J., \& {Richard}, J. 2013, \apj, 763, 9

\bibitem[{{Zahid} {et~al.}(2014){Zahid}, {Kashino}, {Silverman}, {Kewley},
  {Daddi}, {Renzini}, {Rodighiero}, {Nagao}, {Arimoto}, {Sanders},
  {Kartaltepe}, {Lilly}, {Maier}, {Geller}, {Capak}, {Carollo}, {Chu},
  {Hasinger}, {Ilbert}, {Kajisawa}, {Koekemoer}, {Kovacs}, {Le F{\`e}vre},
  {Masters}, {McCracken}, {Onodera}, {Scoville}, {Strazzullo}, {Sugiyama},
  {Taniguchi}, \& {COSMOS Team}}]{Zahid:2014om}
{Zahid}, H.~J., {Kashino}, D., {Silverman}, J.~D., {et~al.} 2014, \apj, 792, 75

\bibitem[{{Zhang} {et~al.}(2018){Zhang}, {Romano}, {Ivison}, {Papadopoulos}, \&
  {Matteucci}}]{Zhang:2018mc}
{Zhang}, Z.-Y., {Romano}, D., {Ivison}, R.~J., {Papadopoulos}, P.~P., \&
  {Matteucci}, F. 2018, \nat, 558, 260

\end{thebibliography}

\appendix 
\label{sec:fake}

\section*{Possibility of fake correlation for the $R_{1213}$-sSFR relation}

\begin{figure*}[h]
\begin{center}
\includegraphics[width=120mm, bb=0 0 979 948]{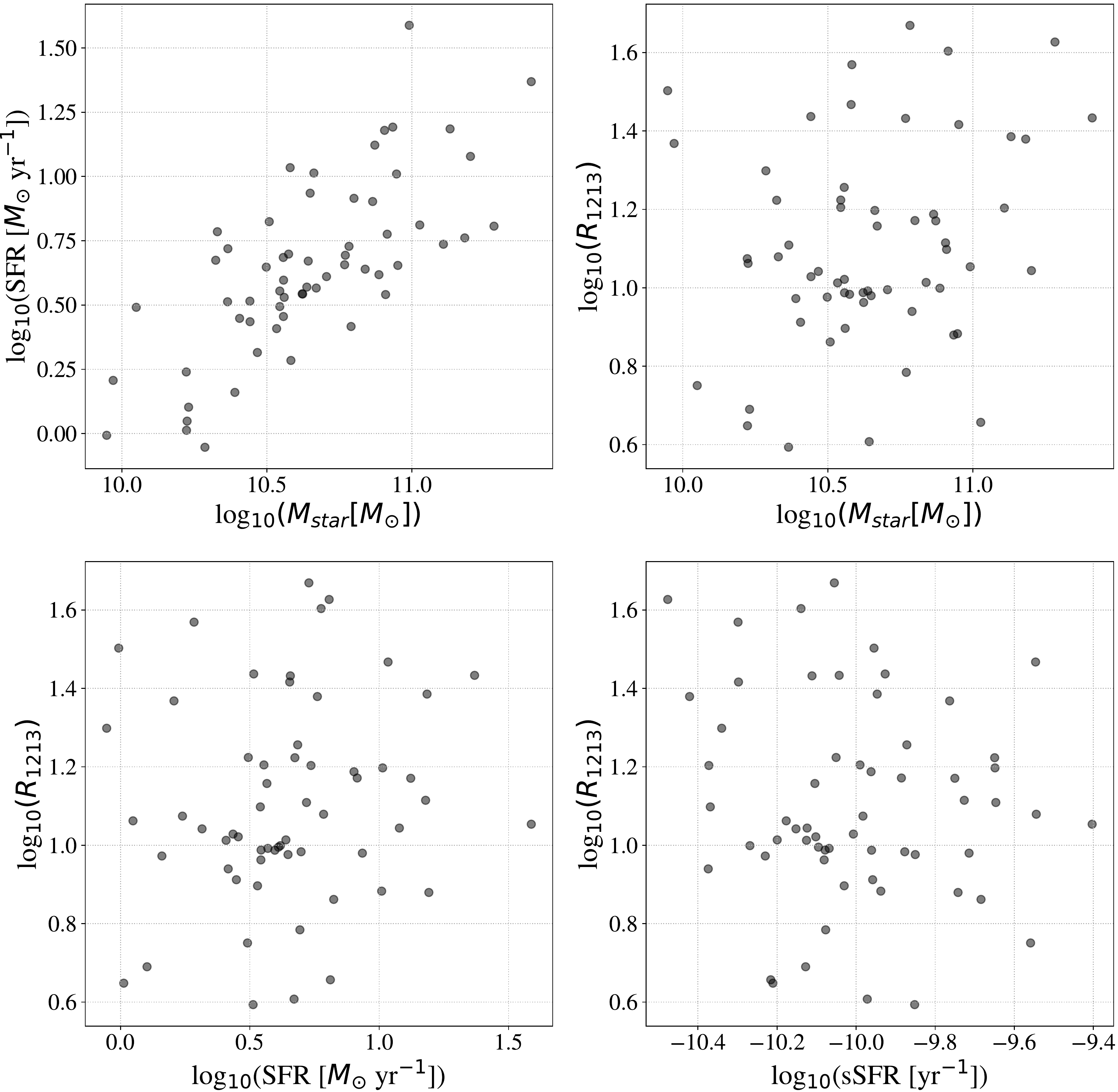}
\end{center}
\caption{
Example of the three-parameter set of the generated 60 mock galaxies and resultant $R_{1213}$-sSFR relation.
}
\label{fig:mock}
\end{figure*}

\begin{figure*}[h]
\begin{center}
\includegraphics[width=\textwidth, bb=0 0 1415 499]{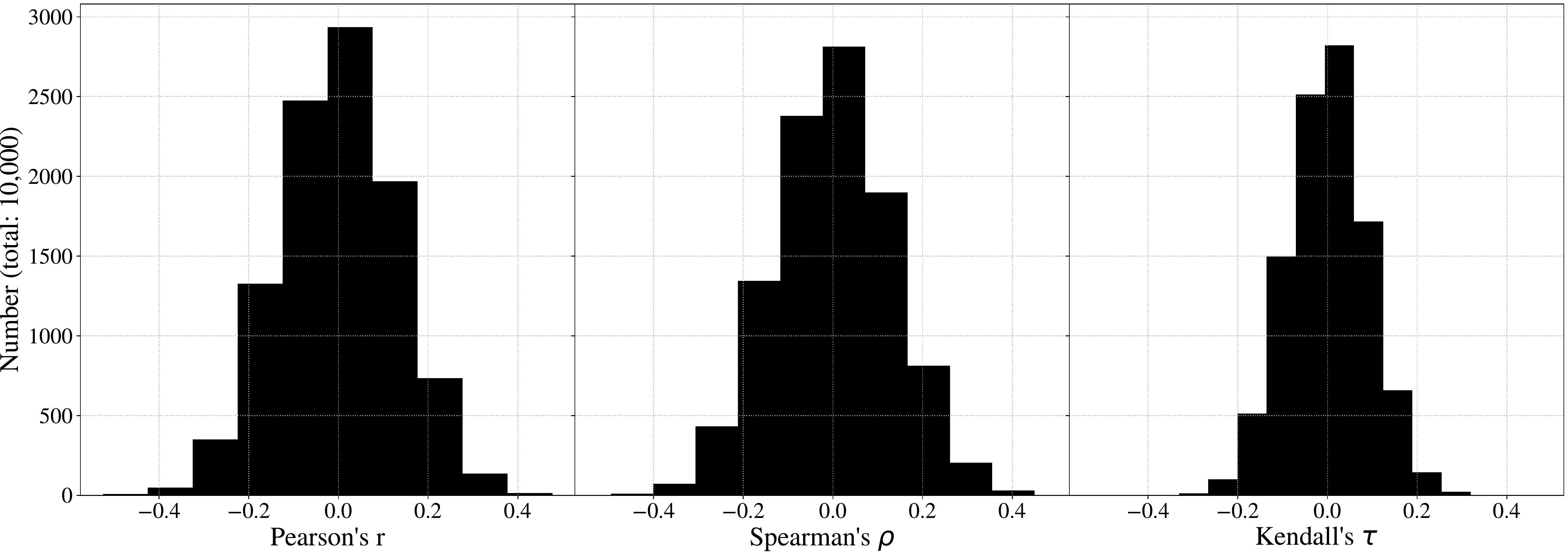}
\end{center}
\caption{
Histograms of the resultant correlation coefficients for each method when generating mock samples for 10,000 times.
}
\label{fig:corrcoef_hist}
\end{figure*}

We find a moderate correlation between $R_{1213}$ and sSFR for the 63 galaxies but no correlation for the $R_{1213}$-SFR (a numerator of sSFR) and $R_{1213}$-$M_{\rm star}$ (a denominator of sSFR) relations in this study.
In order to examine whether or not the obtained correlation of the $R_{1213}$-sSFR relation is fake due to the small number of galaxies, we generate 10,000 sets of mock galaxy samples based on the mean vector and the variance-covariance matrix obtained from the observed $M_{\rm star}$, SFR, and $R_{1213}$ values of the COMING galaxies using the {\tt Multivariate\_normal} function of the {\tt Python} module {\tt Numpy}.
For the number of galaxies in each set, three cases are considered : 60, 100, and 1000.
Then we manually set the covariances of the $R_{1213}$-$M_{\rm star}$ and $R_{1213}$-SFR relations to be zero.
The relationship among these three values of a mock galaxy sample is shown in figure~\ref{fig:mock}.
Using this 10,000 mock data sets, we calculate sSFR and correlation coefficients of the resultant $R_{1213}$-sSFR relation.

Figure~\ref{fig:corrcoef_hist} shows the histogram of the obtained values of the Pearson's r, Spearman's $\rho$, and Kendall's $\tau$ for the 10,000 mock data sets.
The largest $r$, $\rho$, and $\tau$ are 0.48, 0.45, and 0.32, respectively.
The number of the mock dataset with larger correlation coefficients than the observed values are 5 for the Pearson's r, 0 for the Spearman's $\rho$, and 3 for the Kendall's $\tau$, i.e., $p$-values of $3\times10^{-4}$, $0\times10^{-4}$, and $1\times10^{-4}$.
We also find that the histograms become sharper for the larger number of galaxies in each data set.
This result shows that the probability of fake correlation reduces as the number of galaxy sample increases.
The sample number of 63 in this study is large enough to conclude that the obtained positive correlation between $R_{1213}$ and sSFR with correlation coefficients $\sim0.4$ is not fake.

\end{document}